\documentclass[12pt]{article}
\usepackage[T1]{fontenc}
\usepackage{xcolor}
\usepackage{amsmath,amsfonts,amsthm,amssymb}
\usepackage{mathtools}
\usepackage{bm}
\usepackage{algorithm}
\usepackage{algpseudocode}
\usepackage{longtable}
\usepackage{caption}
\usepackage{subcaption}

\title{Parameter estimation of the Heston volatility model with jumps in the asset prices}
\author{Jarosław Gruszka, Janusz Szwabiński}
\date{}
\begin{document}

\maketitle


\begin{abstract}
    Parametric estimation of stochastic differential equations (SDEs) has been a subject of intense studies already for several decades. The Heston model for instance is driven by two coupled SDEs and is often used in financial mathematics for the dynamics of the asset prices and their volatility. Calibrating it to real data would be very useful in many practical scenarios. It is very challenging however, since the volatility is not directly observable. In this paper, a complete estimation procedure of the Heston model without and with jumps in the asset prices is presented. Bayesian regression combined with the particle filtering method is used as the estimation framework. Within the framework, we propose a novel approach to handle jumps in order to neutralise their negative impact on the estimates of the key parameters of the model. An improvement of the sampling in the particle filtering method is discussed as well. Our analysis is supported by numerical simulations of the Heston model to investigate the performance of the estimators. And a practical follow-along recipe is given to allow for finding adequate estimates from any given data.
\end{abstract}

\section{Introduction}

The problem of parameter estimation of mathematical models applied in the fields of economy and finance is of critical importance. In order to use most of the models, like the ones for pricing financial instruments or finding an optimal investment portfolio, one needs to provide values of the model parameters which are often not easily available. For example, a famous, Nobel-prize winning Black-Scholes model for pricing European options \cite{black_pricing_1973} assumes that the dynamics of the underlying asset is what we now call the Geometric Brownian Motion (GBM in short) --- a stochastic process, which has two parameters commonly called the drift and the volatility. Knowing the values of those parameters for a particular underlying instrument is required to make use of the model, as they need to be plugged into the formulas the model provides\footnote{to be very precise, in the context of pricing derivative instruments, in which the Black-Scholes model is usually applied, only the value of the volatility parameter is really required, as the drift variable vanishes during the procedure of changing the probability measure to the risk-neutral one, but the principle argument that knowing values of the model parameters is crucial still holds.}.

Over the last decades, mathematical models describing the behaviour of observed market quantities (e.g. prices of assets, interest rates etc.) became more and more complicated to be able to reflect some particular characteristics of their dynamics. For instance, the phenomenon called the \textit{volatility smile} is widely observed across various types of options and on different markets, however, it is not possible to "configure" the classical Black-Scholes model to reproduce it \cite{meissner_capturing_2001}. Similarly, financial markets occasionally experience sudden drops in the value of assets traded, which can be treated as discontinuities in their trajectories, yet GBM, as a time-continuous model, would never display any kind of a jump in the value of the modelled asset. Therefore --- a need for more complex models emerges, like the ones of Heston \cite{heston_closed-form_1993} and Bates \cite{bates_jumps_1996}  which were designed to address those two specific issues respectively. The problem is that more complicated models typically use more parameters, which need to be estimated and moreover --- standard estimation techniques, like Maximum Likelihood Estimators (MLE) or Generalised Method of Moments (GMM), fail very often for them~\cite{johannes_chapter_2010}. Apart from that, most existing methods for estimating the parameters of more complex financial models, like the ones of Heston or Bates work in the context of derivative instruments only and as such, they require options prices as an input despite the fact that the models themselves actually describe the dynamics of the underlying instruments. This presents two major problems. The first one is that mentioned models are not always used in the context of derivative instruments, as sometimes we are only interested in modelling stock price dynamics (e.g. in research related to stock portfolio management). The other problem is that historical values of basic instruments like indices, stocks or commodities are much easier to be found publicly on the Internet, compared to the options prices. Thus, from the data availability perspective, the estimation tools based on the values of underlying instrument outcompete the ones which require prices of derivatives as an input.   

There is a wide range of methods which use prices of the actual instruments, instead of prices of the derivatives for parameter estimation but the Bayesian approach \cite{lindley_bayes_1972} seems to be especially effective in that field. Among methods based on Bayesian inference, the ones using Monte Carlo Markov Chains (MCMC) are the most prominent for complex financial models. In this group of methods one assumes some distribution for the value of each of the parameters of a model (called the \textit{prior distribution}) and uses it, along with the data, to produce what is called the \textit{posterior distribution} --- samples from which we can treat as possible values of our parameters (see Ref.~\cite{johannes_chapter_2010} for a great overview of MCMC methods used for financial mathematics).

The MCMC concept can be applied in multiple ways and by utilising various different algorithms, including Gibbs sampling or the Metropolis-Hastings algorithm \cite{chib_understanding_1995}, depending on the complexity of the problem. Both are generally very useful for effective estimation of "single" parameters, i.e. those parameters which only have one, constant number as their value. However, some models assume that the directly observable dynamic quantities (e.g. prices) are dependent on other dynamically changing properties of the model. The latter are often called latent variables or state variables. In case of the Heston model for example, the volatility process is a state variable. Estimation of state variables is inherently more complicated than of the regular parameters, as each value which was observed directly was partly determined by value of the state variable at that particular point of time. A very elegant solution to this complication is a methodology called particle filtering. It is based on the idea of creating a collection of values (called particles) which are meant to represent the distribution of the latent variable at a given point of time. Each particle then has a probability assigned to it, which serves as a measure of how likely it is that a given value of the state variable generated the outcome observed at a given moment of time. For an overview of particle filtering methods, we recommend Refs.~\cite{johannes_optimal_2009}~and~\cite{doucet_tutorial_2009}.

Methods outlined above have been studied quite thoroughly for the past years. However, the research articles and literature focuses on the theoretical aspect of the estimation process and is often lacking precision and concreteness. This is in fact a big issue since applying the results of theoretical research in practice almost always requires estimation in one way or another. In our last paper we were studying performance of various investment portfolios depending on assets they contain, which were represented by trajectories of Heston model \cite{gruszka_advanced_2021}. The behaviour of those portfolios turned out to be dependent on some of the assets' characteristics, which are captured by the values of certain parameters of the Heston model. Hence --- an estimation scheme would allow us to see if a given strategy is suitable for a particular asset portfolio. And this is just one example of how the estimation of a financial market model can be utilised. 

In this paper we present a complete set-up for parameter estimation of the Heston model, using only the prices of the basic instrument one wants to study using the model (an index, a stock, a commodity etc., no derivative prices needed). We provide the estimation process for both the pure Heston model and its extended version, with the inclusion of Merton-style jumps (discontinuities), which is then known as the Bates model. In section \ref{sec:Heston_model}, we present the Heston model as well as its extension allowing for the appearance of jumps. We also present a way of changing the time-character of the model from a continuous in time to a discrete one. In section \ref{sec:Estimation_framework}, we describe in details the posterior distributions which one can sample from to obtain the parameters of the model. We also provide a detailed description of the particle filtering scheme needed to reconstruct the volatility process. The whole procedure is summarised in an easy-to-follow pseudo-code algorithm. An exemplary estimation as well as the analysis of the factors that impact the quality of the estimation in general is presented in section \ref{sec:Estimation_results_analysis}. Finally, some conclusions are drawn in the last section.

\section{Heston model --- without and with jumps}
\label{sec:Heston_model}
\subsection{Model characterisation}

Heston model can be described using two stochastic differential equations, one for the process of prices and one for the process of volatility \cite{heston_closed-form_1993}

\begin{gather}
	\label{eq:Heston_stock}
	dS(t) = \mu S(t) dt + \sqrt{v(t)} S(t) d B^S(t),\\
	\label{eq:Heston_vol}
	dv(t) = \kappa(\theta - v(t))dt + \sigma \sqrt{v(t)} S(t) d B^v(t),
\end{gather}
where $t \in [0, T]$. In equation \eqref{eq:Heston_stock}, the parameter $\mu$ represents the drift of the stock price. Equation \eqref{eq:Heston_vol} is widely known as the CIR (i.e. Cox-Ingersoll-Ross) model, featuring an interesting quality called mean-reversion \cite{cox_theory_1985}. Parameter $\theta$ is the long-term average from which the volatility diverges and to which it then comes back, $\kappa$ is the rate of those fluctuations (the bigger $\kappa$, the longer it takes to come back to $\theta$).  Parameter $\sigma$ is called the volatility-of-the-volatility and it is generally responsible for the "scale" of randomness of the volatility process. 

Both stochastic processes are based on their respective Brownian motions --- $B^s(t)$ and $B^v(t)$. Heston model allows for the possibility of those two processes being correlated with an instantaneous correlation coefficient $\rho$,

\begin{equation}
	\label{eq:Heston_rho}
	dB^S(t) dB^v(t) = \rho dt.
\end{equation}

To complete the set-up, deterministic initial conditions for $S$ and $v$ need to be specified. 

\begin{gather}
	\label{eq:Heston_ic}
	S(0) = S_0 > 0,\\
	v(0) = v_0 >0.
\end{gather}

The trajectories coming form the Heston model are continuous, although the model itself can easily be extended to include discontinuities. The most common type of jumps which can easily be incorporated into the model is called Merton log-normal jump. To add it, one needs to augment the equation \eqref{eq:Heston_stock} with an additional term,

\begin{equation}
    \label{eq:Heston_stock_jump}
    dS(t) = \mu S(t) dt + \sqrt{v(t)} S(t) d B^S(t) + (e^{Z(t)}-1) S(t) dq(t),
\end{equation}
where $Z(t)$ is a series of i.i.d. normally distributed random variables with mean $\mu^J$ and standard deviation $\sigma^J$, whereas $q(t)$ is a Poisson counting process with constant intensity $\lambda$. The added term turns the Heston model into Bates model \cite{bates_jumps_1996}. The above extension has an easy real-life interpretation. Namely, $e^{Z(t)}$ is the actual (absolute) rate of the difference between the price before the jump at time $t$ and right after it, i.e. $S(t-)\cdot e^{Z(t)} = S(t+)$. So if for example, for a given $t$, $e^{Z(t)} \approx 0.85$, that means the stock experienced $\sim 15\%$ drop in value at that moment.

\subsection{Euler-Maruyama discretisation}

In order to make the model  applicable in practice, one needs to discretise it, that is --- to rewrite the continuous (theoretical) equations in such a way that the values of the process are given in specific, equidistant points of time. To this end, we split the domain $[0, T]$, into $n$ short intervals, each of length $\Delta t$. Thus $n\cdot\Delta t = T$. To properly transform the SDE's of the model into this new time domain, a discretisation scheme is necessary. We used Euler-Maruyama discretisation scheme for that purpose \cite{kloeden_numerical_1992}. The stock price equation \eqref{eq:Heston_stock} can be discretised as 

\begin{multline}
	\label{eq:Heston_s_disc}
	S(k\Delta t) - S\Big((k-1)\Delta t\Big) = \mu S\Big((k-1)\Delta t\Big)\Delta t + \\
	S\Big((k-1)\Delta t\Big) \sqrt{v\Big((k-1)\Delta t\Big)} \varepsilon ^S(k\Delta t) \sqrt{\Delta t},
\end{multline}
where $k \in \{1, \cdots, n\}$ and $\varepsilon^S$ is a series of $n$ i.i.d. standard normal random variables. 

To highlight the ratio between two consecutive values of the stock price, Eq.~\eqref{eq:Heston_s_disc} is often re-written as
\begin{equation}
	\label{eq:Heston_s_disc_ret}
	\frac{S(k\Delta t)}{S\Big((k-1)\Delta t\Big)} = \mu \Delta t +1  + \sqrt{v\Big((k-1)\Delta t\Big)} \varepsilon ^S(k\Delta t) \sqrt{\Delta t}.
\end{equation}

The same discretisation scheme can be applied to the Eq.~\eqref{eq:Heston_vol}, to obtain:

\begin{multline}
	\label{eq:Heston_v_disc}
	v(k\Delta t) - v\Big((k-1)\Delta t\Big) = \kappa\Bigg(\theta - v\Big((k-1)\Delta t\Big)\Bigg)\Delta t + \\
	\sigma \sqrt{v\Big((k-1)\Delta t\Big)} \varepsilon ^v (k\Delta t) \sqrt{\Delta t}.
\end{multline}

If $\rho = 0$, then $\varepsilon^v$ in the above formula  is also a series of $n$ i.i.d. standard normal random variables. However, if $\rho \neq 0$, then --- to ensure the proper dependency between $S$ and $v$ --- we take 

\begin{equation}
	\label{eq:eps_v}
	\varepsilon ^v (k\Delta t) = \rho \varepsilon ^S(k\Delta t) + \sqrt{1-\rho^2} \varepsilon^{add} (k\Delta t)
\end{equation}
where $\varepsilon^{add}$ is an additional series of $n$ i.i.d. standard normal random variables, which are "mixed" with the ones from $\varepsilon^S$ and hence --- become dependent on them.

\section{Estimation framework}
\label{sec:Estimation_framework}
Estimation of the Heston model consists of two major parts. First one is estimating parameters of the model, i.e. $\mu$, $\kappa$, $\theta$, $\sigma$ and $\rho$, for the basic version of the model and additionally $\lambda$, $\mu^J$ and $\sigma^J$ after inclusion of jumps. The second part is estimating the state variable --- volatility $v(t)$. For all the estimation procedures which we will present here we used the Bayesian inference methodology, in particular Monte Carlo Markov Chains (for parameter estimation within the base model) and particle filtering (for estimation of the volatility as well as jump-related parameters).

\subsection{Regular Heston model}

In order to estimate the Heston model with no jumps, we will mainly be using the principles of Bayesian inference, and in particular --- Bayesian regression ~\cite{ohagan_kendalls_1994}. 

\subsubsection{Estimation of $\mu$}

We will start by finding a way to estimate the drift parameter $\mu$. First, Eq.~\eqref{eq:Heston_s_disc_ret} will be transformed to a regression form. To this end, we will introduce several additional variables. The first, $\eta$, is defined as
\begin{equation}
	\label{eq:eta}
	\eta = \mu\Delta t + 1.
\end{equation}

Let $R(t)$ be a series  of ratios between consecutive prices of assets,
\begin{equation}
	\label{eq:Returns}
	R(k\Delta t) = \frac{S(k\Delta t)}{S\Big((k-1)\Delta t)\Big)},
\end{equation}
for $k\in\{1,2,\ldots, n\}$. Taking the above definitions into consideration, Eq.~\eqref{eq:Heston_s_disc_ret} can be rewritten as:
	
\begin{equation}
	\label{eq:Heston_s_est}
	R(k\Delta t) = \eta + \sqrt{v\Big((k-1)\Delta t\Big)} \varepsilon ^S (k\Delta t) \sqrt{\Delta t}.
\end{equation}

Now, let us divide both sides of this equation by $\sqrt{v\Big((k-1)\Delta t\Big)}\sqrt{\Delta t}$, as $\Delta t$ is known and, at this stage, we consider $v(t)$ to be known too. Let us now introduce another two new variables --- $y^S(t)$ as

\begin{equation}
	\label{eq:y_s}
	y^S(k\Delta t) = \frac{1}{\sqrt{v\Big((k-1)\Delta t\Big)}\sqrt{\Delta t}}R(k\Delta t)
\end{equation}
and $x^S(t)$ as

\begin{equation}
	\label{eq:x_s}
	x^S(k\Delta t) = \frac{1}{\sqrt{v\Big((k-1)\Delta t\Big)}\sqrt{\Delta t}}.
\end{equation}
Inserting them into Eq.~\eqref{eq:Heston_s_est} gives
\begin{equation}
	\label{eq:Heston_s_est2}
	y^S(k\Delta t) = \eta x^S(k\Delta t)  + \varepsilon ^S(k\Delta t). 
\end{equation}

The last expression has the form of a linear regression with  $y^S(t)$ explained by $x^S(t)$. We want to treat it with the Bayesian regression framework. To this end we first collect all discretised values of $y^S(t)$ and $x^S(t)$ into $n$-element column vectors --- $\mathbf{y}^S$ and $\mathbf{x}^S$ respectively,

\begin{equation}
	\label{eq:y_s_vec}
	\mathbf{y}^S = \frac{1}{\sqrt{\Delta t}}
	\begin{bmatrix}
		\frac{R(\Delta t)}{\sqrt{v(0)}} &
		\frac{R(2 \Delta t)}{\sqrt{v(\Delta t)}} & 
		\ldots &
		\frac{R(n\Delta t)}{\sqrt{v\big((n-1)\Delta t\big)}}
	\end{bmatrix}^\prime ,
\end{equation}
\begin{equation}
	\label{eq:x_s_vec}
	\mathbf{x}^S = \frac{1}{\sqrt{\Delta t}}
	\begin{bmatrix}
		 \frac{1}{\sqrt{v(0)}} &
		 \frac{1}{\sqrt{v(\Delta t)}} & 
		 \ldots &
		 \frac{1}{\sqrt{v\big((n-1)\Delta t\big)}}
	\end{bmatrix} ^\prime ,
\end{equation}
where the prime symbol is used for the transpose.


Assuming a prior distribution for $\eta$ to be normal with mean $\mu_0^\eta$ and standard deviation $\sigma_0^\eta$, it follows from the Bayesian regression general results \cite{ohagan_kendalls_1994} that the posterior distribution for $\eta$ will also be normal with precision (inverse of variance) $\tau^\eta$, which can be calculated as
\begin{equation}
\label{eq:tau_eta}
\tau^\eta = \left(\bm{x}^S\right)' \cdot \bm{x}^S + \tau_0^\eta.
\end{equation}
Here, $\tau_0^\eta$ is precision of the prior distribution, i.e. $\tau_0^\eta= \frac{1}{\left(\sigma_0^\eta\right)^2}$. Mean  $\mu^\eta$ of the posterior distribution is of the following form

\begin{equation}
\label{eq:mu_eta}
\mu^\eta = \frac{1}{\tau^\eta} \left( \tau_0^\eta \mu_0^\eta + \left(\bm{x}^S\right)' \cdot \bm{x}^S \hat{\eta}\right),
\end{equation}
where $\hat{\eta}$ is a classical, ordinary-lest-squares (OLS) estimator of $\eta$, i.e.

\begin{equation}
\label{eq:eta_hat}
\hat{\eta} = \left(\left(\bm{x}^S\right)' \cdot \bm{x}^S\right)^{-1}\left(\bm{x}^S\right)'\bm{y}^S.
\end{equation}
Hence --- we can sample realisations of $\eta$ as follows: 

\begin{equation}
	\label{eq:eta_i_bayes}
	\eta_i \sim \mathcal{N}\left(\mu^\eta, \frac{1}{\sqrt{\tau^\eta}}\right)
\end{equation}
where $i$ indicates the $i$-th sample from the posterior distribution which has been found for $\eta$. Having a realisation of $\eta$ in form of $\eta_i$, we can quickly turn it into a realisation of the $\mu$ parameter itself by a simple transform, inverse to Eq.~\eqref{eq:eta}

\begin{equation}
	\label{eq:mu_i}
	\mu_i = \frac{\eta_i-1}{\Delta t}.
\end{equation}

\subsubsection{Estimation of $\kappa$, $\theta$ and $\sigma$}

In order to estimate parameters related to the volatility process, i.e.  $\kappa$, $\theta$ and $\sigma$, we will do a similar exercise but this time using the volatility process. Let us first rewrite Eq.~\eqref{eq:Heston_v_disc} as

\begin{multline}
\label{eq:Heston_v_disc_2}
v(k\Delta t) = \kappa\theta\Delta t + (1-\kappa\Delta t)v\Big((k-1)\Delta t\Big) + \\ \sigma \sqrt{v\Big((k-1)\Delta t\Big)} \varepsilon ^v (k\Delta t) \sqrt{\Delta t}.
\end{multline}
Now let us introduce two new parameters,
\begin{equation}
	\label{eq:beta_1}
	\beta_1 = \kappa \theta \Delta t
\end{equation}
and
\begin{equation}
	\label{eq:beta_2}
	\beta_2 = 1-\kappa\Delta t. 
\end{equation}
From Eqs.~\eqref{eq:Heston_v_disc_2}--\eqref{eq:beta_2} we get
\begin{equation}
	\label{eq:Heston_v_est}
	v(k\Delta t) = \beta_1 + \beta_2 v\Big((k-1)\Delta t\Big) + \sigma \sqrt{v\Big((k-1)\Delta t\Big)} \varepsilon ^v(k\Delta t) \sqrt{\Delta t}.
\end{equation}
In a fashion similar to the equation for the stock price, we can rewrite this last expression as
\begin{multline}
	\label{eq:Heston_v_est2}
	\frac{v(k\Delta t)}{\sqrt{\Delta t} \sqrt{v\Big((k-1)\Delta t\Big)}} = 
	\frac{\beta_1}{\sqrt{\Delta t}\sqrt{v\Big((k-1)\Delta t\Big)}} + \\
	\frac{\beta_2 v\Big((k-1)\Delta t\Big)}{\sqrt{\Delta t} \sqrt{v\Big((k-1)\Delta t\Big)}} + \sigma \varepsilon^v(k\Delta t).
\end{multline}
Introducing vectors
\begin{equation}
	\label{eq:beta_vec}
	\bm{\beta} = 
	\begin{bmatrix}
		 \beta_1 \\ 
		 \beta_2
	\end{bmatrix},
\end{equation}
\begin{equation}
	\label{eq:y_v_vec}
	\mathbf{y}^v = \frac{1}{\sqrt{\Delta t}}
	\begin{bmatrix}
		 \frac{v(2\Delta t)}{\sqrt{v(\Delta t)}} &
		 \frac{v(3\Delta t)}{\sqrt{v(2\Delta t)}} & 
		 \ldots &
		 \frac{v(n\Delta t)}{\sqrt{v\big((n-1)\Delta t\big)}}
	\end{bmatrix}^\prime,
\end{equation}
\begin{equation}
	\label{eq:x_1_v_vec}
	\mathbf{x}_1^v = \frac{1}{\sqrt{\Delta t}}
	\begin{bmatrix}
		 \frac{1}{\sqrt{v(\Delta t)}} &
		 \frac{1}{\sqrt{v(2\Delta t)}} & 
		 \ldots & 
		 \frac{1}{\sqrt{v\big((n-1)\Delta t\big)}}
	\end{bmatrix}^\prime,
\end{equation}
\begin{multline}
	\label{eq:x_2_v_vec}
	\mathbf{x}_2^v = \frac{1}{\sqrt{\Delta t}}
	\begin{bmatrix}
		 \frac{v(\Delta t)}{\sqrt{v(\Delta t)}} &
		 \frac{v(2\Delta t)}{\sqrt{v(2\Delta t)}} & 
		 \ldots &
		 \frac{v\big((n-1)\Delta t\big)}{\sqrt{v\big((n-1)\Delta t\big)}}
	\end{bmatrix}^\prime
	= \\
	\frac{1}{\sqrt{\Delta t}}
	\begin{bmatrix}
		 \sqrt{v(\Delta t)} &
		 \sqrt{v(2\Delta t)} & 
		 \ldots &
		 \sqrt{v\big((n-1)\Delta t\big)}
	\end{bmatrix}^\prime,
\end{multline}
allows us to rewrite the original volatility equation in form of a linear regression
\begin{equation}
	\label{eq:v_reg}
	\mathbf{y}^v = \mathbf{X}^v\bm{\beta} + \sigma\bm{\varepsilon}^v,
\end{equation}
where 
\begin{equation}
	\label{eq:x_v_vec}
	\mathbf{X}^v = 
	\begin{bmatrix}
		 \mathbf{x}_1^v & \mathbf{x}_2^v
	\end{bmatrix}
\end{equation}
and
\begin{equation}
	\label{eq:eps_v_vec}
	\bm{\varepsilon}^v = 
	\begin{bmatrix}
		 \varepsilon^v(\Delta t) &
		 \varepsilon^v(2\Delta t) &
		 \ldots &
		 \varepsilon^v\big((n-1)\Delta t\big)
	\end{bmatrix}
\end{equation}

Using the formulas for Bayesian regression and assuming multivariate (2-dimensional) normal prior for $\bm{\beta}$  with mean vector $\bm{\mu}_0^\beta$ and precision matrix $\bm{\Lambda}_0^\beta$, we get the conjugate posterior distribution being also multivariate normal with precision matrix given by  

\begin{equation}
	\label{eq:Lambda_beta}
	\bm{\Lambda}^\beta = \left(\bm{X}^v\right)' \cdot \bm{X}^v + \bm{\Lambda}_0^\beta
\end{equation}
and mean vector given by 

\begin{equation}
	\label{eq:mu_beta}
	\bm{\mu}^\beta = \left(\bm{\Lambda}^\beta\right)^{-1} \left( \bm{\Lambda}_0^\beta \bm{\mu}_0^\beta + \left(\bm{X}^v\right)' \cdot \bm{X}^v \hat{\bm{\beta}}\right)
\end{equation}
where, again, $\hat{\bm{\beta}}$ is a standard OLS estimator of $\bm{\beta}$,

\begin{equation}
\label{eq:beta_hat}
\hat{\bm{\beta}} = \left(\left(\bm{X}^v\right)' \cdot \bm{X}^v\right)^{-1}\left(\bm{X}^v\right)'\bm{y}^v.
\end{equation}
We can then use this posterior distribution of $\bm{\beta}$ for sampling

\begin{equation}
	\label{eq:beta_i_bayes}
	\bm{\beta}_i \sim \mathcal{N}(\bm{\mu}^\beta, \sigma_{i-1}^2(\bm{\Lambda}^\beta)^{-1})
\end{equation}

It is worth noting that the realisation of $\sigma$ appears in the equation \eqref{eq:beta_i_bayes}, however, we have not defined it yet. This is because both the distribution of $\bm{\beta}$ is dependent on $\sigma$ and the distribution of $\sigma$ is dependent on $\bm{\beta}$. Hence, we suggest taking the realisation of $\sigma$ from the previous iteration here (which is indicated by the $i-1$ subscript). We will address the order of performing calculations in more details further down the article.

Obtaining realisations of the actual parameters is very easy --- one simply needs to inverse the equations defining $\beta_1$ and $\beta_2$

\begin{equation}
	\label{eq:kappa_i_bayes}
	\kappa_i = \frac{1 - \bm{\beta}_i[2]}{\Delta t}
\end{equation}
and

\begin{equation}
	\label{eq:theta_i_bayes}
	\theta_i = \frac{\bm{\beta}_i[1]}{\kappa_i\Delta t}
\end{equation}
where $\bm{\beta}_i[1]$ and $\bm{\beta}_i[2]$ are respectively the first and the second component of the $\bm{\beta}_i$ vector. 

The most common approach for estimating $\sigma$ is assuming the inverse-gamma prior distribution for $\sigma^2$. If the parameters of the prior distribution are $a_0^\sigma$ and $b_0^\sigma$, then the conjugate posterior distribution is also inverse gamma 

\begin{equation}
\label{eq:sigma_i_bayes}
(\sigma_i)^2 \sim \mathcal{I}\mathcal{G}\left(a^\sigma, b^\sigma\right)
\end{equation}
where 

\begin{equation}
	\label{eq:a_sigma}
	a^\sigma = a_0^\sigma + \frac{n}{2}
\end{equation}
and

\begin{equation}
	\label{eq:b_sigma}
	b^\sigma = b_0^\sigma + \frac{1}{2}\left(\left(\bm{y}^v\right)' \cdot \bm{y}^v 
											 + \left(\bm{\mu}_0^\beta\right)' \bm{\Lambda}_0^\beta \bm{\mu}_0^\beta 
											 - \left(\bm{\mu}^\beta\right)' \bm{\Lambda}^\beta \bm{\mu}^\beta \right).
\end{equation}

\subsubsection{Estimation of $\rho$}

For the estimation of $\rho$ we follow an approach presented in Ref.~\cite{jacquier_bayesian_2004}. We first define the residuals for the stock price equation

\begin{equation}
	\label{eq:e1_rho}
	e_1^\rho(k\Delta t) = \frac{R(k\Delta t) - \mu_i\Delta t - 1}{\sqrt{\Delta t}\sqrt{v\Big((k-1)\Delta t\Big)}}
\end{equation}
and for the volatility equation,

\begin{equation}
	\label{eq:e2_rho}
	e_2^\rho(k\Delta t) = \frac{v(k\Delta t) - v\Big((k-1)\Delta t\Big) - \kappa_i\Bigg(\theta_i -v\Big((k-1)\Delta t\Big)\Bigg)\Delta t}{\sqrt{\Delta t}\sqrt{v\Big((k-1)\Delta t\Big)}}.
\end{equation}

By calculating those residuals we try to retrieve the error terms from the Eqs.~\eqref{eq:Heston_stock} and~\eqref{eq:Heston_vol} --- $\varepsilon^S(t)$ and $\sigma\varepsilon^v(t)$ respectively, as we know they are tied with each other by a relationship given by equation \eqref{eq:eps_v}. Taking this fact into consideration, we end up with the following equation

\begin{equation}
    \label{eq:rho_reg}
   e_2^\rho(k\Delta t) = \sigma\rho e_1^\rho (k\Delta t) + \sigma\sqrt{1-\rho^2} \varepsilon^{add} (k\Delta t) 
\end{equation}
We now introduce two new variables, traditionally called $\psi = \sigma\rho$ and $\omega = \sigma^2(1-\rho^2)$. It is not difficult to deduce that the relationship between $\rho$ and a newly-introduced variables $\psi$ and $\omega$ is 

\begin{equation}
	\label{eq:rho}
	\rho = \frac{\psi}{\sqrt{\psi^2 + \omega}}
\end{equation}
Then, Eq.~\eqref{eq:rho_reg} becomes 

\begin{equation}
    \label{eq:rho_reg2}
    e_2^\rho(k\Delta t) = \psi e_1^\rho (k\Delta t) + \sqrt{\omega} \varepsilon^{add} (k\Delta t),
\end{equation}
which is again, a linear regression of $e_2^\rho(t)$ on $e_1^\rho$. Thus, we can use the exact same estimation scheme as in case of previously described regressions. We first collect the values of  $e_1^\rho$ and $e_2^\rho$ in two $n$-element vectors:

\begin{equation}
	\label{eq:e1_rho_vec}
	\mathbf{e}_1^\rho =
	\begin{bmatrix}
		 e_1^\rho(\Delta t) & 
		 e_1^\rho(2\Delta t) & 
		 \ldots & 
		 e_1^\rho\left(n\Delta t\right)
	\end{bmatrix}^\prime,
\end{equation}

\begin{equation}
	\label{eq:e2_rho_vec}
	\mathbf{e}_2^\rho =
	\begin{bmatrix}
		 e_2^\rho(\Delta t) & 
		 e_2^\rho(2\Delta t) & 
		 \ldots & 
		 e_2^\rho\left(n\Delta t\right)
	\end{bmatrix}^\prime.
\end{equation}
Then we appose both vectors, forming into an $n$-by-2 matrix:

\begin{equation}
	\label{eq:e_rho}
	\mathbf{e}^\rho = 
	\begin{bmatrix}
		 \mathbf{e}_1^\rho & \mathbf{e}_2^\rho
	\end{bmatrix}
\end{equation}
Next, we define a $2$-by-$2$ matrix $\mathbf{A}^\rho $ as 

\begin{equation}
	\label{eq:A_rho}
	\mathbf{A}^\rho = (\mathbf{e}^\rho)' \cdot \mathbf{e}^\rho 
\end{equation}
If we assume a normal prior for $\psi$ with mean $\mu_0^\psi$ and precision $\tau_0^\psi$, the posterior distribution for $\psi$ is going to also be normal with mean $\mu^\psi$ given by

\begin{equation}
	\label{eq:mu_psi}
	\mu^\psi = \frac{\mathbf{A}_{12}^\rho + \mu_0^\psi\tau_0^\psi}{\mathbf{A}_{11}^\rho + \tau_0^\psi}
\end{equation}
and precision $\tau^\psi$ equal to

\begin{equation}
	\label{eq:tau_psi}
	\tau^\psi = \mathbf{A}_{11}^\rho + \tau_0^\psi,
\end{equation}
where $\mathbf{A}_{11}^\rho$, $\mathbf{A}_{12}^\rho$ and $\mathbf{A}_{22}^\rho$ are the elements of the matrix $\mathbf{A}^\rho$ on positions $(1,1)$, $(1,2)$ and $(2,2)$ respectively. 

Assuming the inverse gamma prior with parameters $a_0^\omega$ and $b_0^\omega$ for $\omega$, the conjugate posterior distribution is also inverse gamma with parameters 

\begin{equation}
	\label{eq:a_psi}
	a^\omega = a_0^\omega + \frac{n}{2}
\end{equation}
and 

\begin{equation}
	\label{eq:b_psi}
	b^\omega = b_0^\omega + \frac{1}{2}\left(\mathbf{A}_{22}^\rho - \frac{(\mathbf{A}_{12}^\rho)^2}{\mathbf{A}_{11}^\rho}\right).
\end{equation}

Thus, sampling from the posterior distribution of $\omega$ can be summarised as 

\begin{equation}
	\label{eq:omega_i_bayes}
	\omega_i \sim \mathcal{I}\mathcal{G}\left(a^\omega, b^\omega\right)
\end{equation}
while when it comes to sampling from $\psi$ it is

\begin{equation}
	\label{eq:psi_i_bayes}
	\psi_i \sim \mathcal{N}\left(\mu^\psi, \frac{\sqrt{\omega_i}}{\sqrt{\tau^\psi}}\right).
\end{equation}

To get $\rho$, we simply make use of Eq.~ \eqref{eq:rho}.

\subsubsection{Estimation of $v(t)$ -- particle filtering}

For all estimation procedures shown in the previous sections, we assumed $v(t)$ to be known. However, in practice, the volatility is not a directly observable quantity, it is "hidden" in the process of prices, which we have access to. Hence --- we need a way to extract volatility from the price process and particle filtering methodology is extremely useful for that purpose. Here, we will only sketch the outline of the particle filtering logic, namely the SIR algorithm, which we utilise to get the volatility estimator. For more in-depth review of particle filtering we suggest the works of Refs.~\cite{johannes_optimal_2009} and \cite{doucet_tutorial_2009}. Here we follow a procedure similar to the one presented in Ref~\cite{christoffersen_volatility_2007}.

We start by fixing the number of particles $N$. In each moment of time $t = k\Delta t$, we will produce $N$ particles, which are going to represent various possible values of the volatility at that point of time. By averaging out all of those particles we will get an estimate of the true volatility $v(t)$. The process of creating the particles is as follows: at the time $t=0$, we create $N$ initial particles, all with the initial value of the volatility, which we assume to be the long term average $\theta$. Denoting each of the particles by $V_j$, for $j \in \{1, 2, \ldots N\}$, we have  

\begin{equation}
	\label{eq:V_0_pf}
	V_j(0) = \theta_i.
\end{equation}

For any subsequent moment of time, except the last one $t = k\Delta t, k\notin\{0, n\}$, we define three sequences of size $N$. $\varepsilon_j$ will be a series of independent, standard normal random variables

\begin{equation}
	\label{eq:eps_pf}
	\varepsilon_j(k\Delta t) \sim \mathcal{N}\left(0,1\right).
\end{equation}
The series $z_j$ contains residuals from the stock price process, where the past values of volatility are replaced by the values of the particles from the previous time step

\begin{equation}
	\label{eq:z_eps_pf}
	z_j(k\Delta t)  = \frac{R(k\Delta t) - \mu_i\Delta t -1}{\sqrt{\Delta t}\sqrt{V_j\Big((k-1)\Delta t\Big)}}.
\end{equation}
And finally the series  $w_j$, which incorporates the possible dependency between the stock process and the volatility particles

\begin{equation}
	\label{eq:w_eps_pf}
	w_j(k\Delta t)  = z_j(k\Delta t)\rho_i + \varepsilon_j(k\Delta t)\sqrt{1-(\rho_i)^2}.
\end{equation}

Having all that, the candidates for the new particles $\widetilde{V}_j$ are created as follows

\begin{multline}
	\label{eq:V_tilde_pf}
	\widetilde{V}_j(k\Delta t)  = V_j\Big((k-1)\Delta t\Big) + \kappa_i\Bigg(\theta_i - V_j\Big((k-1)\Delta t\Big)\Bigg)\Delta t + \\
	\sigma_i \sqrt{\Delta t}\sqrt{V_j\Big((k-1)\Delta t\Big)} w_j.
\end{multline}
Each candidate for a particle is evaluated based on how probable it is that such value of the volatility would generate the return that has actually been observed. The measure of this probability $\widetilde{W}_j$ is a value of a normal distribution PDF function designed specifically for this purpose\footnote{Equation \eqref{eq:W_tilde_pf} is the reason we cannot run this procedure for $k=n$, as we would not be able to obtain $R\big((n+1)\Delta t\big)$, since the last available value is $R(n\Delta t)$.}

\begin{equation}
	\label{eq:W_tilde_pf}
	\widetilde{W}_j(k\Delta t) = 
	    \frac{1}{\sqrt{2\pi\widetilde{V}_j(k\Delta t)\Delta t}} 
	    \exp{\left(-\frac{1}{2} \frac{\Big(R\big((k+1)\Delta t\big) - \mu_i\Delta t -1\Big)^2}{\widetilde{V}_j(k\Delta t) \Delta t}\right)}.
\end{equation}

To be able to treat the values of the proposed measure along with the values of particles as a proper probability distribution on its own, we normalise them, so that their sum is equal to 1,

\begin{equation}
	\label{eq:W_bow_pf}
	\breve{W}_j(k\Delta t) = \widetilde{W}_j(k\Delta t)\left(\sum_{j=1}^N{\widetilde{W}_j(k\Delta t)}\right)^{-1}.
\end{equation}

Now, we combine the particles with and their respective probabilities, forming two-element vectors $\mathbf{U}_j$

\begin{equation}
	\label{eq:U_pf}
	\mathbf{U}_j(k\Delta t) = \big(\widetilde{V}_j(k\Delta t), \breve{W}_j(k\Delta t)\big).
\end{equation}

We would now want to sample from the probability distribution described by $\mathbf{U}_j$ to get the true, "refined" particles. Most sources suggest drawing from it, treating it as a multinomial distribution. However, this makes all the "refined" particles having the same values as the "raw" ones --- just the proportions are changed (the same "raw" particle can be drawn several times, if it has probability bigger than others). To address this problem, we will do the sampling in a different way. We first need to sort the values of particles in an ascending order. Mathematically speaking, we create another sequence and call it $\widetilde{V}_j^{sort}$, ensuring that the following conditions are all met:
\begin{enumerate}
    \item the particle with the smallest value will be the first in the new sequence, i.e. 
        \begin{equation}
	        \label{eq:V_sorted_1_pf}
	        \widetilde{V}_1^{sort}(k\Delta t) = \min_{j\in\{1,2,\ldots,N\}}{\{\widetilde{V}_j(k\Delta t)\}},
        \end{equation}
    \item the particle with the biggest value will be the last in the new sequence, i.e.
        \begin{equation}
	        \label{eq:V_sorted_2_pf}
	        \widetilde{V}_N^{sort}(k\Delta t) = \max_{j\in\{1,2,\ldots,N\}}{\{\widetilde{V}_j(k\Delta t)\}},
        \end{equation}
    \item for any $j \in \{2, 3, \ldots N-1\}$ we will have
        \begin{equation}
	        \label{eq:V_sorted_3_pf}
        	\widetilde{V}_{j-1}^{sort}(k\Delta t)<\widetilde{V}_{j}^{sort}(k\Delta t) < \widetilde{V}_{j+1}^{sort}(k\Delta t).
    \end{equation}
\end{enumerate}

We also want to keep track of the probabilities of our sorted particles, so we order the probabilities in the very same way, by defining another probability sequence $\breve{W}_j^{sort}$, 

\begin{equation}
	\label{eq:W_bow_sorted_pf}
	\breve{W}_j^{sort}(k\Delta t) = \breve{W}_m(k\Delta t) \text{ for } m \text{ such that } \widetilde{V}_{j}^{sort}(k\Delta t) \in \mathbf{U}_m(k\Delta t)
\end{equation}

We will now make $\widetilde{V}_1^{sort}$ and $\widetilde{V}_N^{sort}$ the edges of the support of our new continuous distribution that we will be sampling from. The CDF function of the distribution is given by the formula below (time labels have been dropped for the sake of legibility, as all variables are evaluated at $t=k\Delta t$):

\begin{equation}
	\label{eq:F_V_pf}
	F_{\widetilde{V}^{sort}}(v) = 
	\begin{cases}
	    0 
	    & \mbox{if } v \leqslant \widetilde{V}_1^{sort}\\
	    
	    \frac{v-\widetilde{V}_1^{sort}}{\widetilde{V}_2^{sort}-\widetilde{V}_1^{sort}}
        (\breve{W}_1^{sort}+\frac{1}{2}\breve{W}_2^{sort}) 
        &\mbox{if } v \in (\widetilde{V}_1^{sort}, \widetilde{V}_{2}^{sort}]\\
        
        \begin{aligned}[b]
        &\left(\sum_{m=1}^{j-1}{\breve{W}_m^{sort}} + \frac{1}{2}\breve{W}_{j}^{sort}\right) + \\
        &\frac{v-\widetilde{V}_j^{sort}}{\widetilde{V}_{j+1}^{sort}-\widetilde{V}_j^{sort}}\left(\frac{1}{2}\breve{W}_j^{sort}+\frac{1}{2}\breve{W}_{j+1}^{sort}\right) \end{aligned}
        & \mbox{if } v \in (\widetilde{V}_j^{sort}, \widetilde{V}_{j+1}^{sort}] \\
        & \mbox{for } j \in \{2, 3, \ldots N-2\}\\
        
        \begin{aligned}[b]
        &\left(\sum_{m=1}^{N-2}{\breve{W}_m^{sort}} + \frac{1}{2}\breve{W}_{N-1}^{sort}\right) + \\
        &\frac{v-\widetilde{V}_{N-1}^{sort}}{\widetilde{V}_{N}^{sort}-\widetilde{V}_{N-1}^{sort}}\left(\frac{1}{2}\breve{W}_{N-1}^{sort}+\breve{W}_{N}^{sort}\right)
        \end{aligned}
        & \mbox{if } v \in (\widetilde{V}_{N-1}^{sort}, \widetilde{V}_{N}^{sort}]\\
        
        1 & \mbox{if } v > \widetilde{V}_N^{sort}
    \end{cases}.
\end{equation}

\begin{figure}[ht]
\centering
\includegraphics[width=\textwidth]{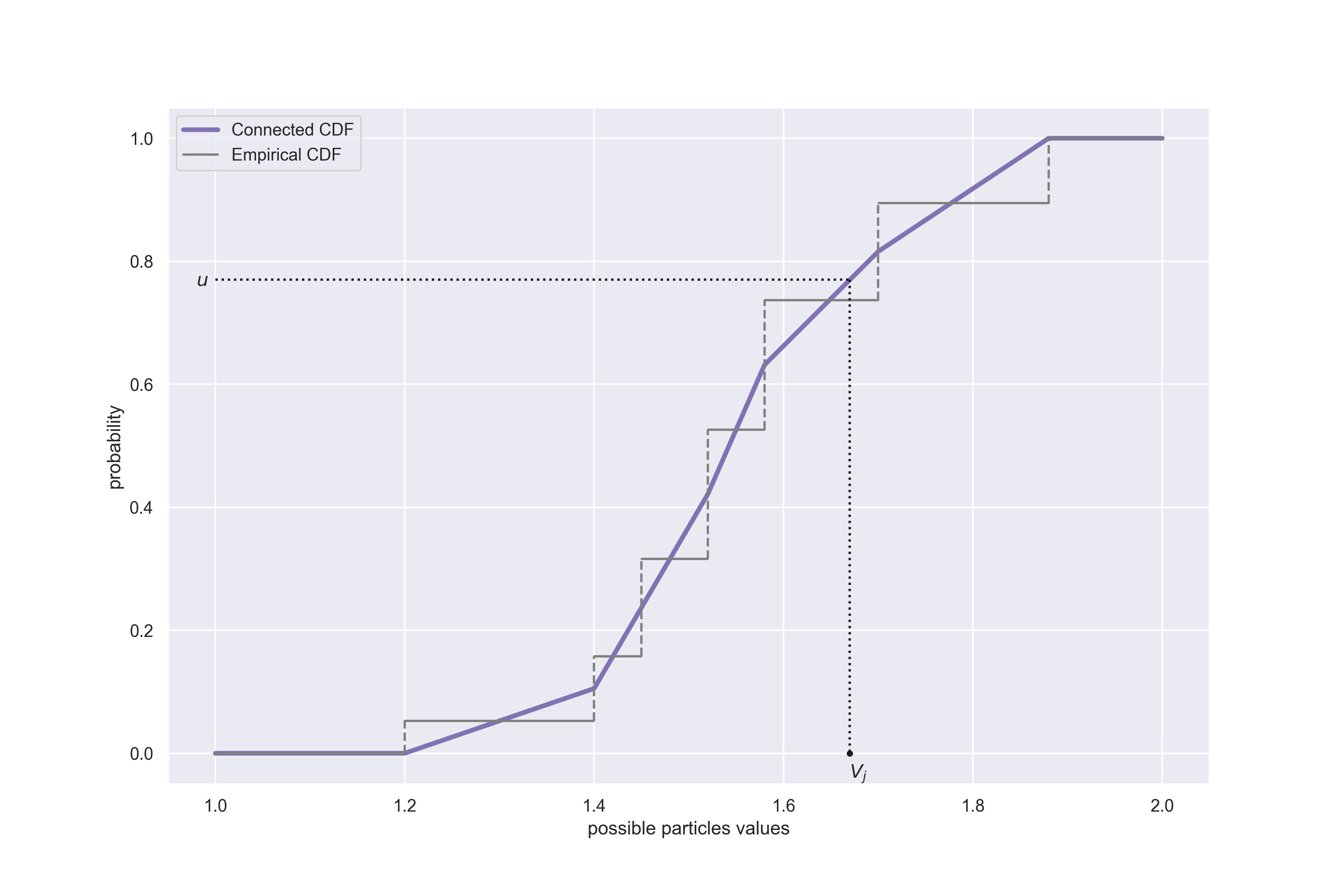}
\caption{Visualisation of the process of resampling particles according to their probabilities. Values of raw particles are the places, where the empirical cumulative ditribution function (ECDF in short) jumps and each jump size represents probability of a respective raw particle. The Connected CDF is a continuous modification of the ECDF, build according to the formula \eqref{eq:F_V_pf}. In order to resample, a uniform random variable $u$ is generated and then its inverse through the Connected CDF function becomes a new, resampled particle --- $V_j$.}
\label{fig:particle_resampling}
\end{figure}
The formula might look overwhelming, but there is a very easy-to-follow interpretation behind it (see Fig.~\ref{fig:particle_resampling}). The new, "refined" particles can be generated by drawing from the distribution given by $F_{\widetilde{V}^{sort}}$ --- simplest way to do it is to use the inverse transform sampling. 

\begin{equation}
	\label{eq:V_pf}
	V_{j}(k\Delta t) \sim F_{\widetilde{V}^{sort}}
\end{equation}
After following the described procedure for each $k \in \{1,2, \ldots, n-1\}$, we can specify the actual estimate of the volatility process as the mean of "refined" particles. 

\begin{equation}
	\label{eq:v_pf}
	v(k\Delta t) = \frac{1}{N} \sum_{j=1}^N{V_{j}(k\Delta t)}.
\end{equation}
For $k = n$, we can simply assume $v(n\Delta t) = v\Big((n-1)\Delta t\Big)$, which should not have any tangible negative impact on any procedures using the $v(t)$ estimate for sufficiently dense time discretisation grid. 

\subsection{Heston model with jumps}

The above estimation framework can be used with minor changes to also estimate Heston model with jumps. The model's SDE is defined in Eq.~\eqref{eq:Heston_stock_jump}. After incorporation of jumps, changes are needed particularly in the particle filtering part of the estimation procedure. The  particles need to be created not only for various possible values of volatility $V_j(t)$, but also for a possibility of a jump in that particular moment of time --- $J_j(t)$ --- and size of that jump --- $Z_j(t)$. So one can now think of a particle as of a 3-element "tuple" $(V_j, J_j, Z_j)$. Generating "raw" values for $J_j$ and $Z_j$ is easy --- for each $j\in \{0,1,\ldots, N\}$, $J_j$ is simply a random variable form a Bernoulli distribution with parameter $\lambda^{th}$,    

\begin{equation}
	\label{eq:J_tilde_pf}
	\widetilde{J}_j(k\Delta t)\sim \mathcal{B}(\lambda^{th}).
\end{equation}
Parameter $\lambda^{th}\in[0, 1)$ can be thought of as a "threshold" value --- a proportion of the number of particles which encode the occurrence of a jump to all the particles. If the number of jumps is expected to be significant, it is good to increase the value of $\lambda^{th}$, hence increasing the number of particles suggesting the jump in each step.

Raw particles for $Z_j$ are simply independent normal random variables with mean $\mu_0^J$ and standard deviation $\sigma_0^J$, which depict our "prior" beliefs about the size and variance of the jumps

\begin{equation}
	\label{eq:Z_tilde_pf}
	\widetilde{Z}_j(k\Delta t)\sim \mathcal{N}(\mu_0^J, \sigma_0^J).
\end{equation}

Assigning probabilities to the particles is different as well, since the normal PDF function which we use is different when there is a jump. Hence, equation \eqref{eq:W_tilde_pf} needs to be updated to 

\begin{equation}
	\label{eq:W_tilde_pf_jump}
	\widetilde{W}_j(k\Delta t) = 
	\begin{cases}
	    \begin{aligned}[b]
	    &\frac{1}{\sqrt{2\pi\widetilde{V}_j\big((k-1)\Delta t\big)\Delta t}} \times \\
	    &\exp{\left(-\frac{1}{2} \frac{\big(R(k\Delta t) - \mu_i\Delta t -1\big)^2}{\widetilde{V}_j\big((k-1)\Delta t\big) \Delta t}\right)} 
	    \end{aligned}
	    &\mbox{if } \widetilde{J}_j =0
        \\
        
        \begin{aligned}[b]
        &\frac{1}{\exp{\big(\widetilde{Z}_j(k\Delta t)\big)}\sqrt{2\pi\widetilde{V}_j\big((k-1)\Delta t\big)\Delta t}}
	    \times \\
	    &\exp{\left(-\frac{1}{2} \frac{\Big(R(k\Delta t) - \exp{\big(\widetilde{Z}_j(k\Delta t)\big)}(\mu_i\Delta t +1)\Big)^2}{\exp{\big(2\widetilde{Z}_j(k\Delta t)\big)}\widetilde{V}_j\big((k-1)\Delta t\big) \Delta t}\right)}  
	    \end{aligned}
	    &\mbox{if } \widetilde{J}_j = 1 
	    
    \end{cases}
\end{equation}

We then normalise $\widetilde{W}_j$ so that it sums to 1 and resample $\widetilde{V}_j$ as in the case with no jumps, but additionally, we  resample $\widetilde{Z}_j$ in the exact same way as $\widetilde{V}_j$, i.e. we sort the particles and draw from the distribution $F_{\widetilde{Z}^{sort}}$ to obtain the "refined" particles $Z_j$,

\begin{equation}
	\label{eq:Z_pf}
	Z_{j}(k\Delta t) \sim F_{\widetilde{Z}^{sort}}.
\end{equation}

Finally, for the estimate of $\lambda$, for each $k\in \{1, 2, \ldots n\}$ one needs to sum the cumulative value of all particles declaring a jump. That way we will get a probability that a jump took place at the time $t=k\Delta t$,

\begin{equation}
	\label{eq:lambda_k}
	\lambda(k\Delta t)=  \sum_{j=1}^{N} J_j(k\Delta t)\breve{W}_j(k\Delta t).
\end{equation}

To get the actual estimate of $\lambda$ one needs to average $\lambda(t)$ across all time points obtained for different values of $k$,

\begin{equation}
	\label{eq:lambda_i}
	\lambda_i =  \frac{1}{T}\sum_{k=1}^{n} \lambda (k\Delta t).
\end{equation}

Similarly, to obtain the estimate of $\mu^J$ and $\sigma^J$, for each $k$ one needs to first calculate the average size of a jump from the refined particles

\begin{equation}
	\label{eq:Z_k_pf}
	Z(k\Delta t) = \frac{1}{N} \sum_{j=1}^N{Z_{j}(k\Delta t)}
\end{equation}
and then calculate the mean and standard deviation of the results, weighed by the probability of a jump at time moment $t$ indicated by $\lambda(t)$. For the weighted mean of the jumps we get

\begin{equation}
	\label{eq:mu_J_i}
	\mu_i^J =  \left(\sum_{k=1}^{n} Z(k\Delta t) \lambda(k\Delta t)\right)\left(\sum_{k=1}^{n}{\lambda(k\Delta t)}\right)^{-1}
\end{equation}
and for the standard deviation:
\begin{equation}
	\label{eq:sigma_J_i}
	\sigma_i^J =  \sqrt{\left(\sum_{k=1}^{n}{\lambda(k\Delta t)\big(Z(k\Delta t) - \mu_i^J\big)^2}\right)\left(\frac{n-1}{n}\sum_{k=1}^{n}{\lambda(k\Delta t)}\right)^{-1}}.
\end{equation}

\begin{figure}[ht]
\centering
\includegraphics[width=\textwidth]{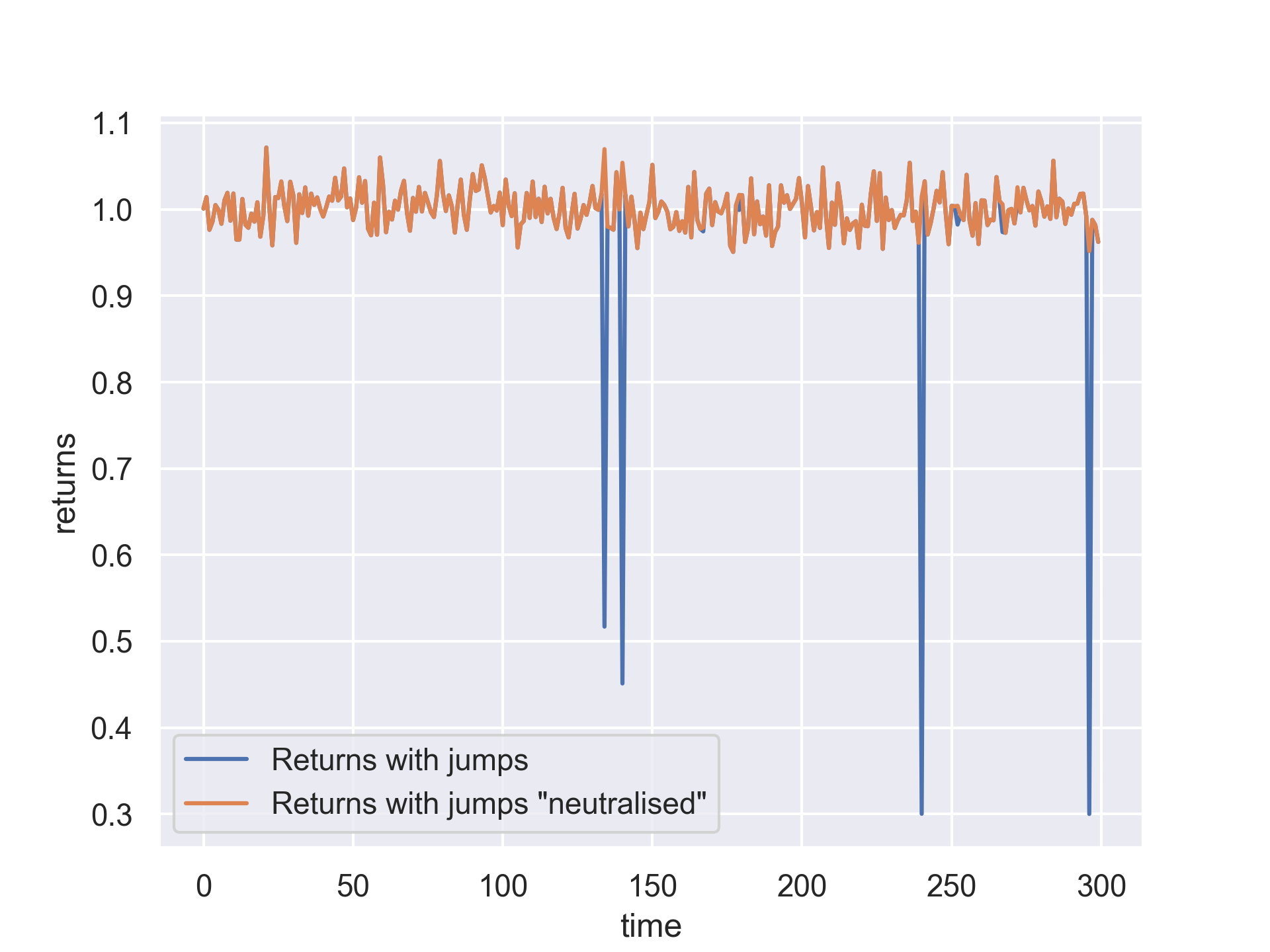}
\caption{Comparison of returns for a process with jumps calculated based on formula \eqref{eq:Returns} (blue line) and \eqref{eq:Returns_jumps} (orange line). It can be clearly seen that the jumps have been "neutralised" in the latter case.}
\label{fig:returns_neurtalised}
\end{figure}

The presence of jumps also influences the estimation of other parameters ---  some of the procedures presented in the previous subsection are not fully correct, as jumps added to the stock price will additionally increase or --- more likely --- decrease the returns. To improve that, a correction of the definitions of $R(t)$ is needed in order to "neutralize" the impact of jumps on the parameters. In other words, eq. \eqref{eq:Returns} should be replaced with

\begin{equation}
	\label{eq:Returns_jumps}
	R(k\Delta t) = \frac{S(k\Delta t)}{S\Big((k-1)\Delta t)\Big)} \Bigg(1 - \lambda(k\Delta t) \Big(1-\exp\big(-Z(k\Delta t)\big)\Big)\Bigg)
\end{equation}
Note that the added term has a value very close to $1$ when $\lambda(t)$ is close to $0$, which indicates there was no jump at time $t$ --- so the correction to the "original" value of $R(t)$ is very minor. However, if $\lambda(t)$ is close to 1, which means there was a jump --- the value of the term gets close to $\exp\big(-Z(t)\big)$, which is an inverse of the jump factor (with the estimated jumps size $Z(t)$). Multiplying by that inverse brings the value of $R(t)$ to a level such as if there was no jump at time $t$ (see Fig. \ref{fig:returns_neurtalised}) and the estimation of the parameters of the model can be carried out as before.

\subsection{Estimation procedure}

The Bayesian estimation framework presented above relies of several parameters for the prior distributions that cannot be calculated within the procedure itself. They are often referred to as metaparameters. For example, for the estimation of the $\mu$ parameter, values of two metaparameters are required --- $\mu_0^\eta$ and $\tau_0^\eta$ (see Eqs. \eqref{eq:tau_eta} and \eqref{eq:mu_eta}). Their values should reflect our preexisting beliefs regarding the value of the parameter, which we are trying to estimate ---  $\mu$ in this case. Let us say that for a given trajectory of the Heston model process, we assume the value of $\mu$ to be around $0.5$. What values should we than choose for the metaparameters? First of all, we need to note that $\mu_0^\eta$ and $\tau_0^\eta$ are not the parameters of the prior distribution for $\mu$ directly. They are parameters of another random variable, which we introduced to utilise the Bayesian framework --- namely $\eta$. The connection between $\mu$ and $\eta$ is known and given by the equation \eqref{eq:eta}. Hence, if --- as we said --- we assume $\mu$ to be around certain value, then, using this relationship, we can deduce the value of $\eta$. And since the prior distibution of $\eta$ is normal, with mean $\mu_0^\eta$ and variance $\sigma_0^\eta = \frac{1}{\tau_0^\eta}$, we can propose the value of the mean of this distribution to be whatever $\eta$ is for the supposed value of $\mu$. Selecting a value for $\sigma_0^\eta$ (and thus --- $\tau_0^\eta$) is even more equivocal --- it should reflect the level of confidence that we have for picking a mean parameter. That is, if we feel that value we chose for $\mu_0^\eta$ would bring us close to the true $\mu$, we should put a smaller value for $\sigma_0^\eta$. However, if we are not so sure about it, bigger value of $\sigma_0^\eta$ should be used. Similar analysis can be repeated for picking the values of other metaparameters. We need to be aware that prior which is used will always in some way influence the final estimate of a given parameter. More detailed analysis on this topic is given in section~\ref{sec:important_findings}. 

Another problem which emerges when applying Bayesian inference (especially for more complex models) is that estimating one parameter often requires knowing a value of some of the others and vice-versa. Hence --- there is not an obvious way of how to start the whole procedure. One way to address this problem is to come up with the initial guesses for the values of all the parameters (as described in the paragraph above) and use them in the first round of samplings. A well designed MCMC estimation algorithm should bring us closer to the true values of parameters with each new round of samplings. In Algorithms \ref{alg:est_Heston} and \ref{alg:est_Heston_jumps} our procedure for the Heston model without and with jumps is shown, respectively. The meaning of all symbols is briefly summarised in Appendix \ref{app:table_of_symbols}.

\begin{algorithm}
\caption{Estimating Heston model}\label{alg:est_Heston}
\begin{algorithmic}
\Require \\
number of samples: $n_s$\\
time step $\Delta t$ \\
maturity $T$\\
prices $S(k\Delta t)$ for $k\in\{0, 1,\ldots, n\}$, so that $n\Delta t = T$\\
number of particles $N$
initial value of $\mu$: $\mu_0$\\
initial value of $\kappa$: $\kappa_0$\\
initial value of $\theta$: $\theta_0$\\
initial value of $\sigma$: $\sigma_0$\\
initial value of $\rho$: $\rho_0$\\
prior distribution parameters for $\eta$: $\mu_0^\eta$ and $\tau_0^\eta$\\
prior distribution parameters for $\bm\beta$: $\bm{\mu}_0^\beta$ and $\bm{\Lambda}_0^\beta$\\
prior distribution parameters for $\sigma^2$: $a_0^\sigma$ and $b_0^\sigma$\\
prior distribution parameters for $\psi$: $\mu_0^\psi$ and $\tau_0^\psi$\\
prior distribution parameters for $\omega$: $a_0^\omega$ and $b_0^\omega$\\
\Ensure \\
estimate of parameter $\mu$: $\hat{\mu}$\\
estimate of parameter $\kappa$: $\hat{\kappa}$\\
estimate of parameter $\theta$: $\hat{\theta}$\\
estimate of parameter $\sigma$: $\hat{\sigma}$\\
estimate of parameter $\rho$: $\hat{\rho}$\\
estimate of the volatility process: $v(t)$\\

\For{$k = 1 \to n$}
    \State set $R(k\Delta t)$ as shown in eq.\eqref{eq:Returns}
\EndFor
\For{$i = 0 \to n_s$}
    \For{$k = 1 \to n-1$}
        \Comment{particle filtering procedure}
        \For{$j = 1 \to N$}
            \State obtain $V_j(k\Delta t)$ as shown in eq. \eqref{eq:V_0_pf} -- \eqref{eq:V_pf}
        \EndFor
    \State obtain $v(k\Delta t)$ as shown in eq. \eqref{eq:v_pf}
    \EndFor
\algstore{est_Heston}
\end{algorithmic}
\end{algorithm}

\begin{algorithm}
\begin{algorithmic}
\algrestore{est_Heston}
    \State obtain $\mu_i$ as shown in eq. \eqref{eq:Heston_s_est} -- \eqref{eq:mu_i}
    \State obtain $\kappa_i$, $\theta_i$ and $\sigma_i$ as shown in eq. \eqref{eq:beta_1} -- \eqref{eq:b_sigma}
    \State obtain $\rho_i$ as shown in eq. \eqref{eq:e1_rho} -- \eqref{eq:psi_i_bayes}
\EndFor
\State set $\hat{\mu} = \frac{1}{n_s}\sum_{i=1}^{n_s}{\mu_i}$
\State set $\hat{\kappa} = \frac{1}{n_s}\sum_{i=1}^{n_s}{\kappa_i}$
\State set $\hat{\theta} = \frac{1}{n_s}\sum_{i=1}^{n_s}{\theta_i}$
\State set $\hat{\sigma} = \frac{1}{n_s}\sum_{i=1}^{n_s}{\sigma_i}$
\State set $\hat{\rho} = \frac{1}{n_s}\sum_{i=1}^{n_s}{\rho_i}$
\end{algorithmic}
\end{algorithm}

\begin{algorithm}
\caption{Estimating Heston model with jumps}\label{alg:est_Heston_jumps}
\begin{algorithmic}
\Require \\
number of samples: $n_s$\\
time step $\Delta t$ \\
maturity $T$\\
prices $S(k\Delta t)$ for $k\in\{0, 1,\ldots, n\}$, so that $n\Delta t = T$\\
number of particles $N$
initial value of $\mu$: $\mu_0$\\
initial value of $\kappa$: $\kappa_0$\\
initial value of $\theta$: $\theta_0$\\
initial value of $\sigma$: $\sigma_0$\\
initial value of $\rho$: $\rho_0$\\
prior distribution parameters for $\eta$: $\mu_0^\eta$ and $\tau_0^\eta$\\
prior distribution parameters for $\bm\beta$: $\bm{\mu}_0^\beta$ and $\bm{\Lambda}_0^\beta$\\
prior distribution parameters for $\sigma^2$: $a_0^\sigma$ and $b_0^\sigma$\\
prior distribution parameters for $\psi$: $\mu_0^\psi$ and $\tau_0^\psi$\\
prior distribution parameters for $\omega$: $a_0^\omega$ and $b_0^\omega$\\
ratio of particle indicating jumps: $\lambda^{th}$\\
prior distribution parameters for $Z$: $\mu_0^J$ and $\sigma_0^J$
\Ensure \\
estimate of parameter $\mu$: $\hat{\mu}$\\
estimate of parameter $\kappa$: $\hat{\kappa}$\\
estimate of parameter $\theta$: $\hat{\theta}$\\
estimate of parameter $\sigma$: $\hat{\sigma}$\\
estimate of parameter $\rho$: $\hat{\rho}$\\
estimate of the volatility process: $v(t)$\\

\For{$k = 1 \to n$}
    \State set $R(k\Delta t)$ as shown in eq.\eqref{eq:Returns}
\EndFor
\For{$i = 0 \to n_s$}
    \For{$k = 1 \to n$}
        \Comment{particle filtering procedure}
        \For{$j = 1 \to N$}
            \State generate $\widetilde{J}_j(k\Delta t)$ as shown in eq. \eqref{eq:J_tilde_pf}
            \State generate $\widetilde{Z}_j(k\Delta t)$ as shown in eq. \eqref{eq:Z_tilde_pf}
            \State obtain $V_j\big((k-1)\Delta t\big)$ as shown in eq. \eqref{eq:eps_pf} -- \eqref{eq:V_pf} and \eqref{eq:W_tilde_pf_jump}
        \EndFor
\algstore{est_Heston}
\end{algorithmic}
\end{algorithm}

\begin{algorithm}
\begin{algorithmic}
\algrestore{est_Heston}
    \State obtain $v(k\Delta t)$ as shown in eq. \eqref{eq:v_pf} 
    \State obtain $Z(k\Delta t)$ and $\lambda(k\Delta t)$ as shown in eq. \eqref{eq:Z_pf} -- \eqref{eq:lambda_k} and \eqref{eq:Z_k_pf}
    \EndFor
\For{$k = 1 \to n$}
    \State update $R(k\Delta t)$ as shown in eq.\eqref{eq:Returns_jumps}
\EndFor
\State obtain $\mu_i$ as shown in eq. \eqref{eq:Heston_s_est} -- \eqref{eq:mu_i}
\State obtain $\kappa_i$, $\theta_i$ and $\sigma_i$ as shown in eq. \eqref{eq:beta_1} -- \eqref{eq:b_sigma}
\State obtain $\rho_i$ as shown in eq. \eqref{eq:e1_rho} -- \eqref{eq:psi_i_bayes}
\State obtain $\lambda_i$ as shown in eq. \eqref{eq:lambda_i}
\State obtain $\mu_i^J$ as shown in eq. \eqref{eq:mu_J_i}
\State obtain $\sigma_i^J$ as shown in eq. \eqref{eq:sigma_J_i}
\EndFor

\State set $\hat{\mu} = \frac{1}{n_s}\sum_{i=1}^{n_s}{\mu_i}$
\State set $\hat{\kappa} = \frac{1}{n_s}\sum_{i=1}^{n_s}{\kappa_i}$
\State set $\hat{\theta} = \frac{1}{n_s}\sum_{i=1}^{n_s}{\theta_i}$
\State set $\hat{\sigma} = \frac{1}{n_s}\sum_{i=1}^{n_s}{\sigma_i}$
\State set $\hat{\rho} = \frac{1}{n_s}\sum_{i=1}^{n_s}{\rho_i}$
\State set $\hat{\lambda} = \frac{1}{n_s}\sum_{i=1}^{n_s}{\lambda_i}$
\State set $\hat{\mu^J} = \frac{1}{n_s}\sum_{i=1}^{n_s}{\mu_i^J}$
\State set $\hat{\sigma^J} = \frac{1}{n_s}\sum_{i=1}^{n_s}{\sigma_i^J}$

\end{algorithmic}
\end{algorithm}

\section{Analysis of the estimation results}
\label{sec:Estimation_results_analysis}

\subsection{Exemplary estimation}
\begin{figure}
     \centering
     \vspace{-3cm}
     \begin{subfigure}[b]{0.45\textwidth}
         \centering
         \includegraphics[width=\textwidth]{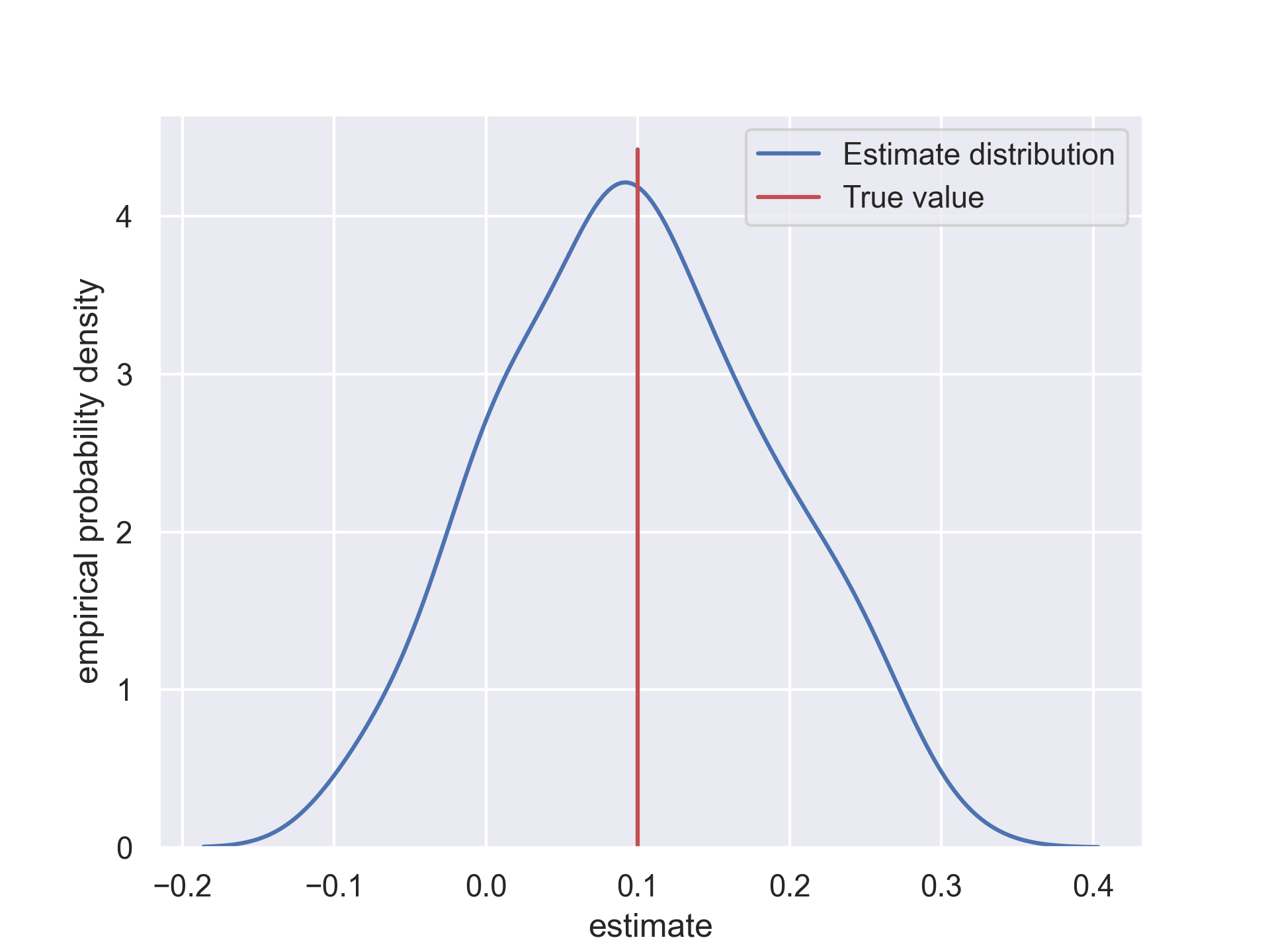}
         \caption{parameter $\mu$}
         \label{fig:exemplary_results_mu}
     \end{subfigure}
     \hfill
     \begin{subfigure}[b]{0.45\textwidth}
         \centering
         \includegraphics[width=\textwidth]{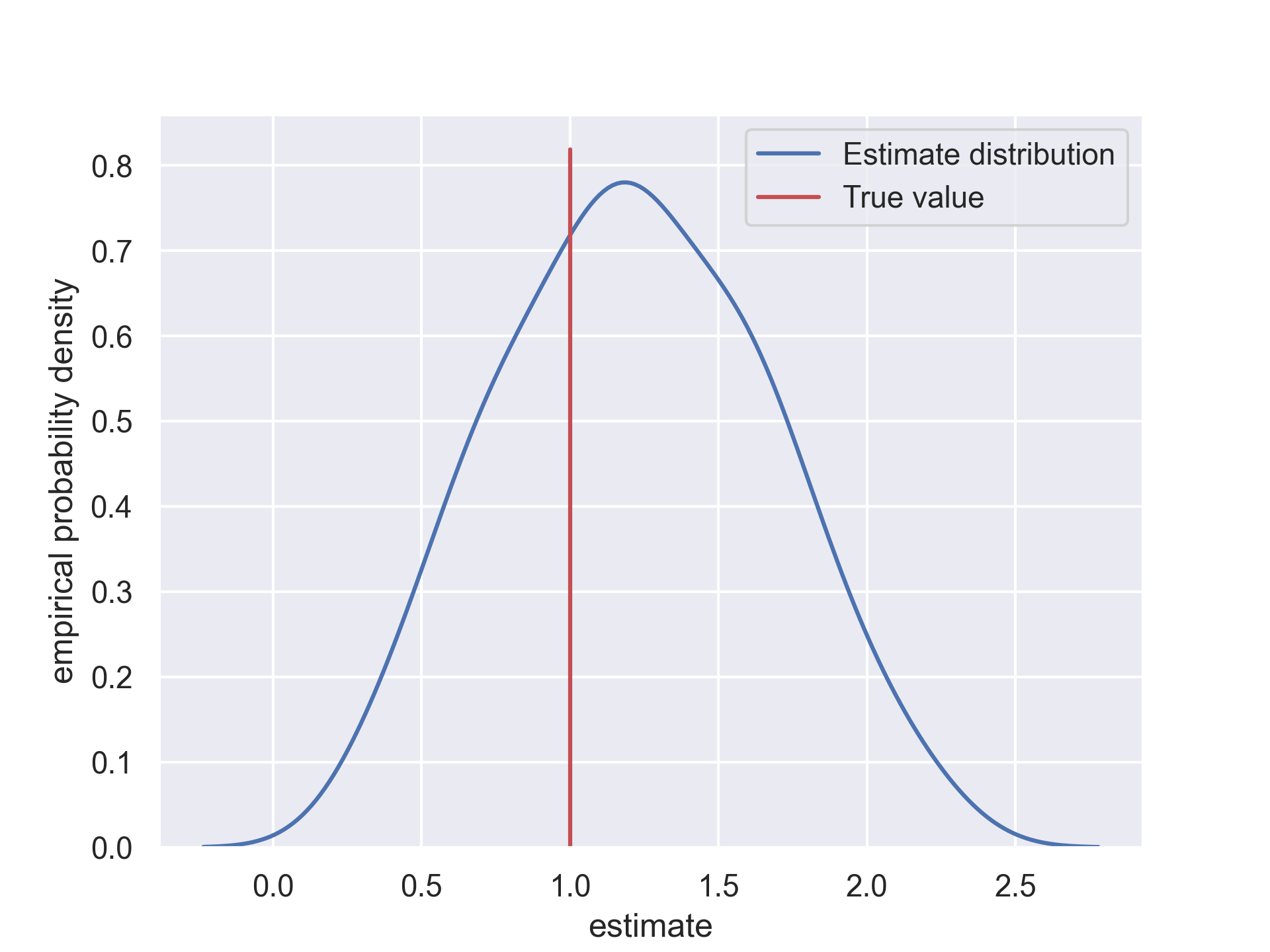}
         \caption{parameter $\kappa$}
         \label{fig:exemplary_results_kappa}
     \end{subfigure}\\
     \begin{subfigure}[b]{0.45\textwidth}
         \centering
         \includegraphics[width=\textwidth]{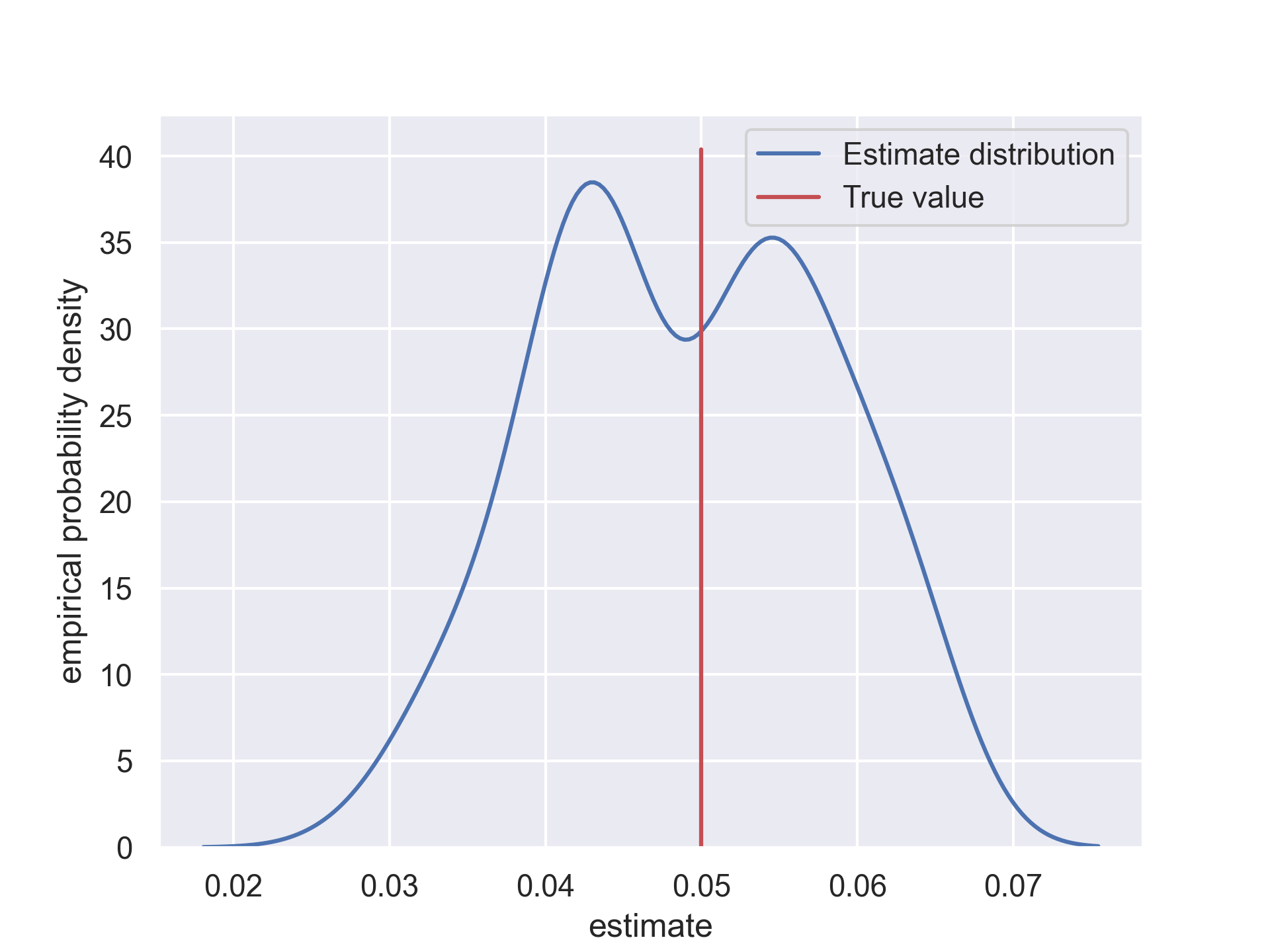}
         \caption{parameter $\theta$}
         \label{fig:exemplary_results_theta}
     \end{subfigure}
     \hfill
     \begin{subfigure}[b]{0.45\textwidth}
         \centering
         \includegraphics[width=\textwidth]{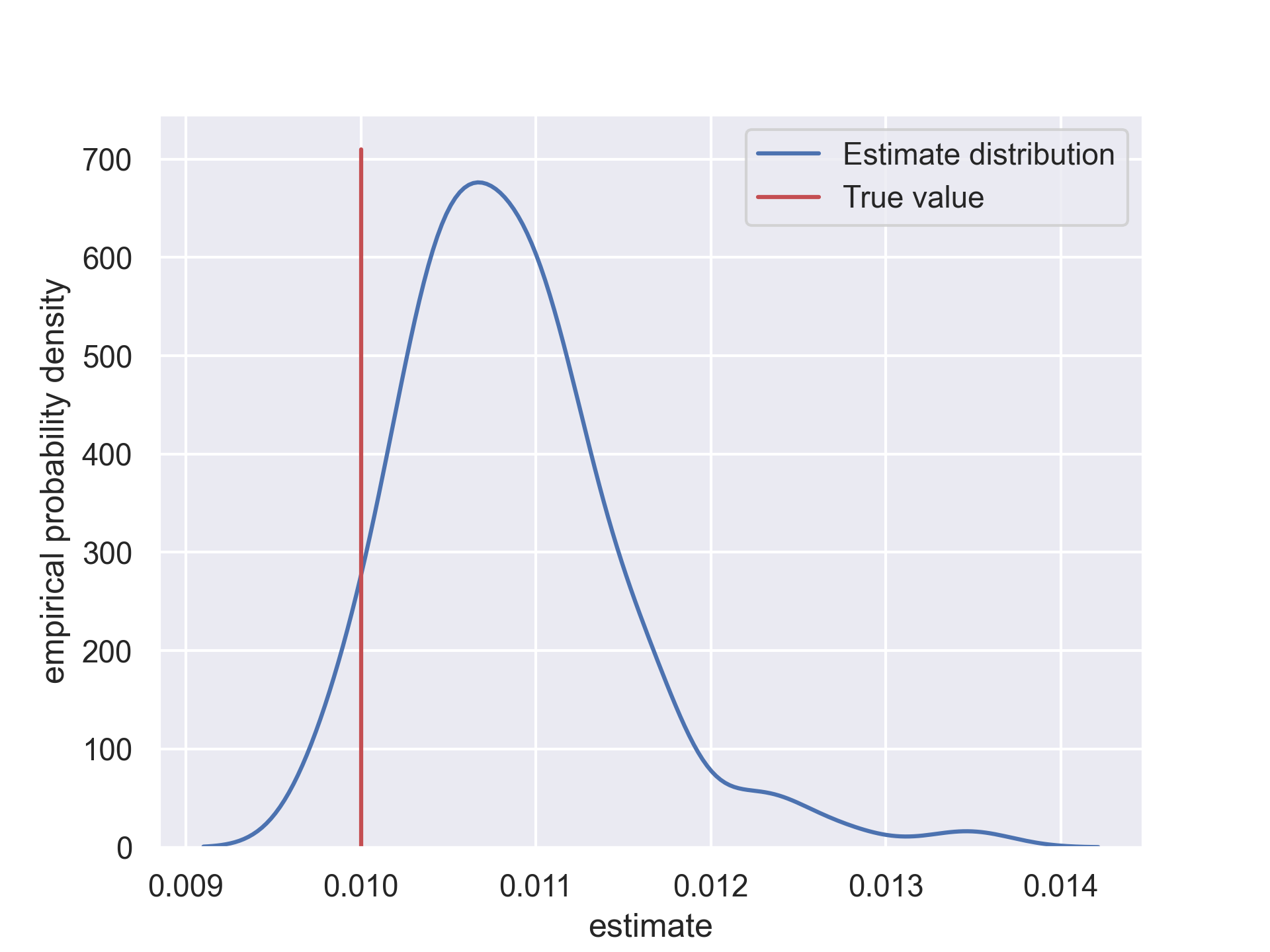}
         \caption{parameter $\sigma$}
         \label{fig:exemplary_results_sigma}
     \end{subfigure}\\
     \begin{subfigure}[b]{0.45\textwidth}
         \centering
         \includegraphics[width=\textwidth]{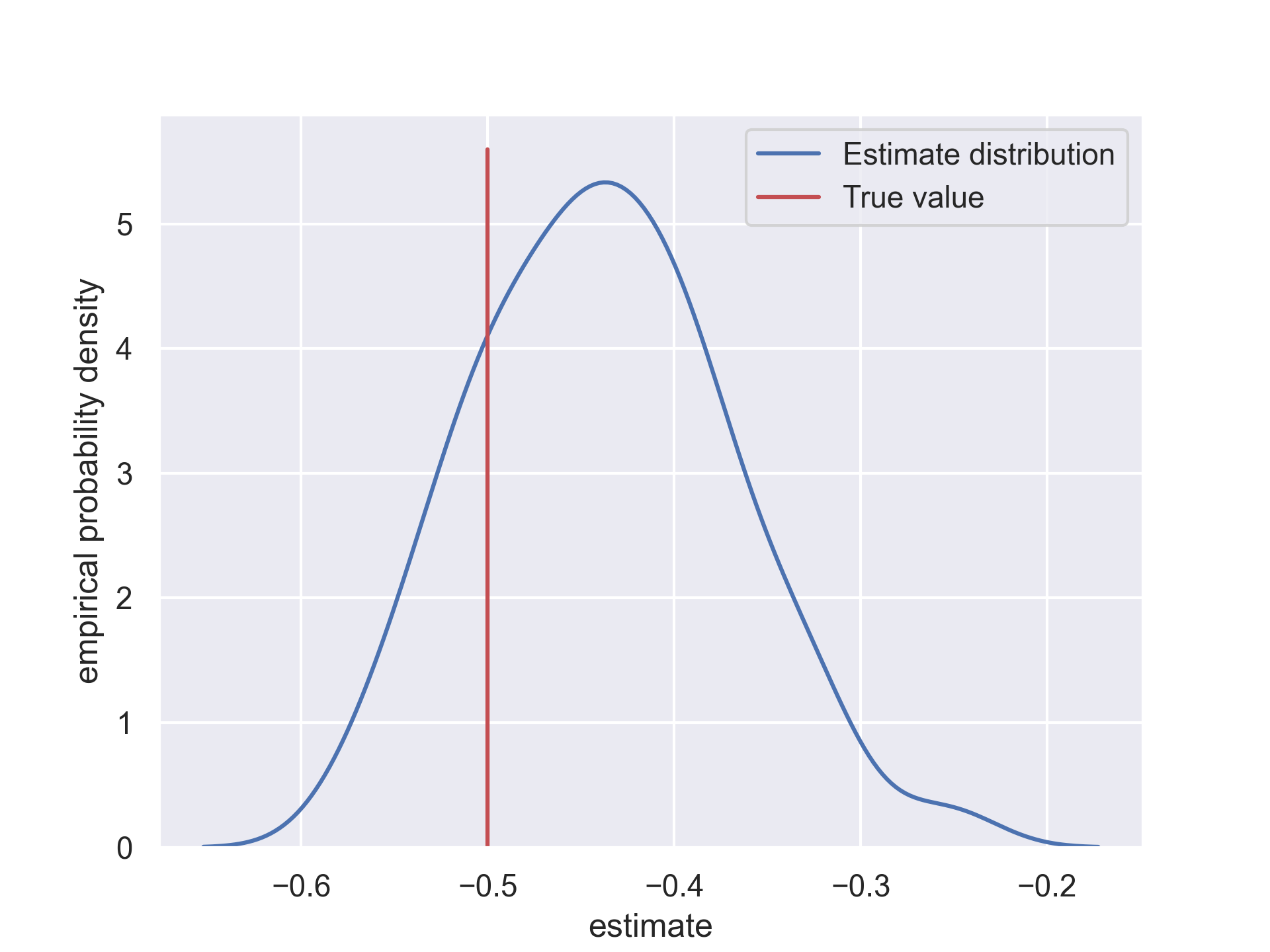}
         \caption{parameter $\rho$}
         \label{fig:exemplary_results_rho}
     \end{subfigure}
     \hfill
     \begin{subfigure}[b]{0.45\textwidth}
         \centering
         \includegraphics[width=\textwidth]{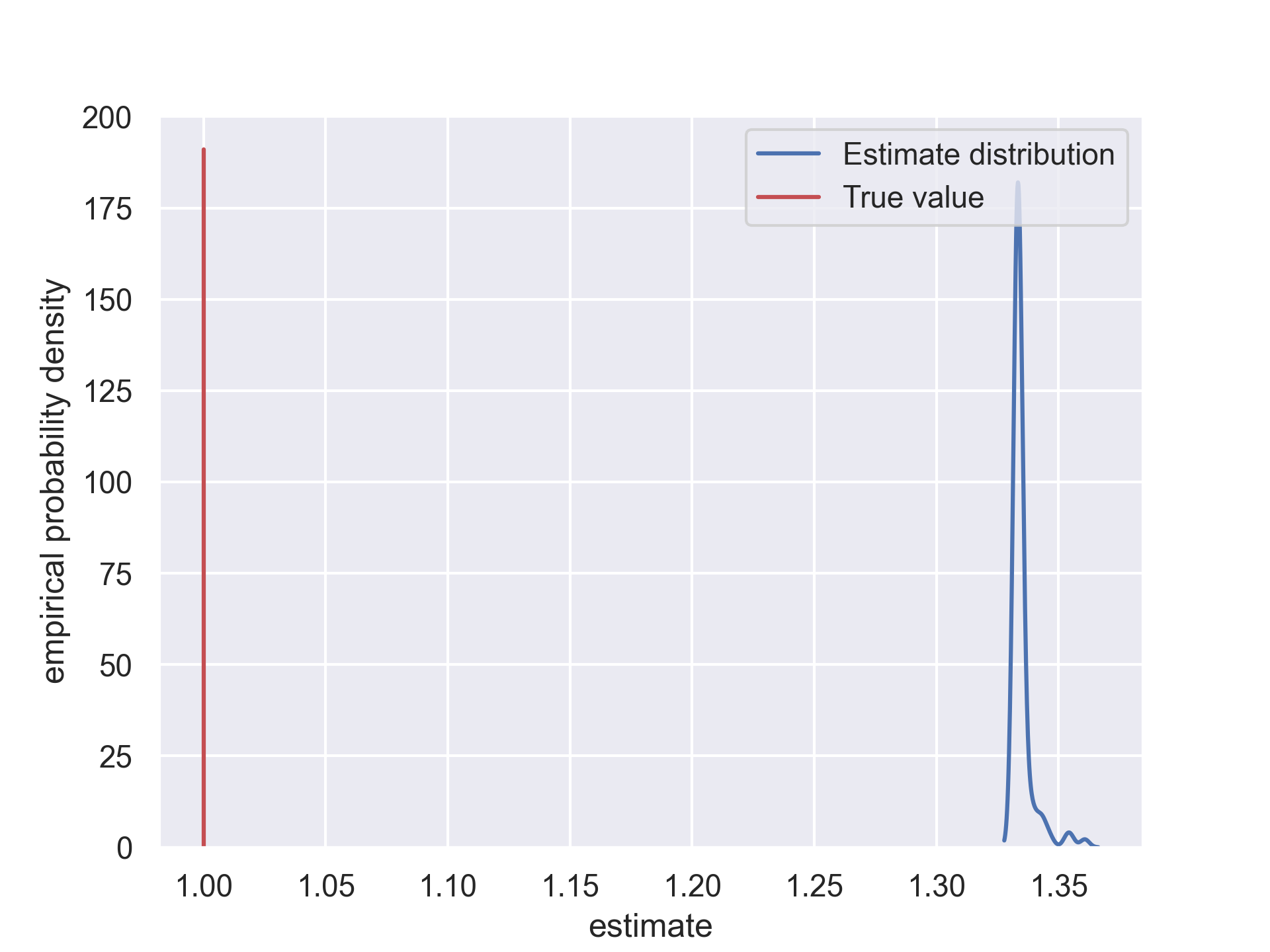}
         \caption{parameter $\lambda$}
         \label{fig:exemplary_results_lambda}
     \end{subfigure}\\
     \begin{subfigure}[b]{0.45\textwidth}
         \centering
         \includegraphics[width=\textwidth]{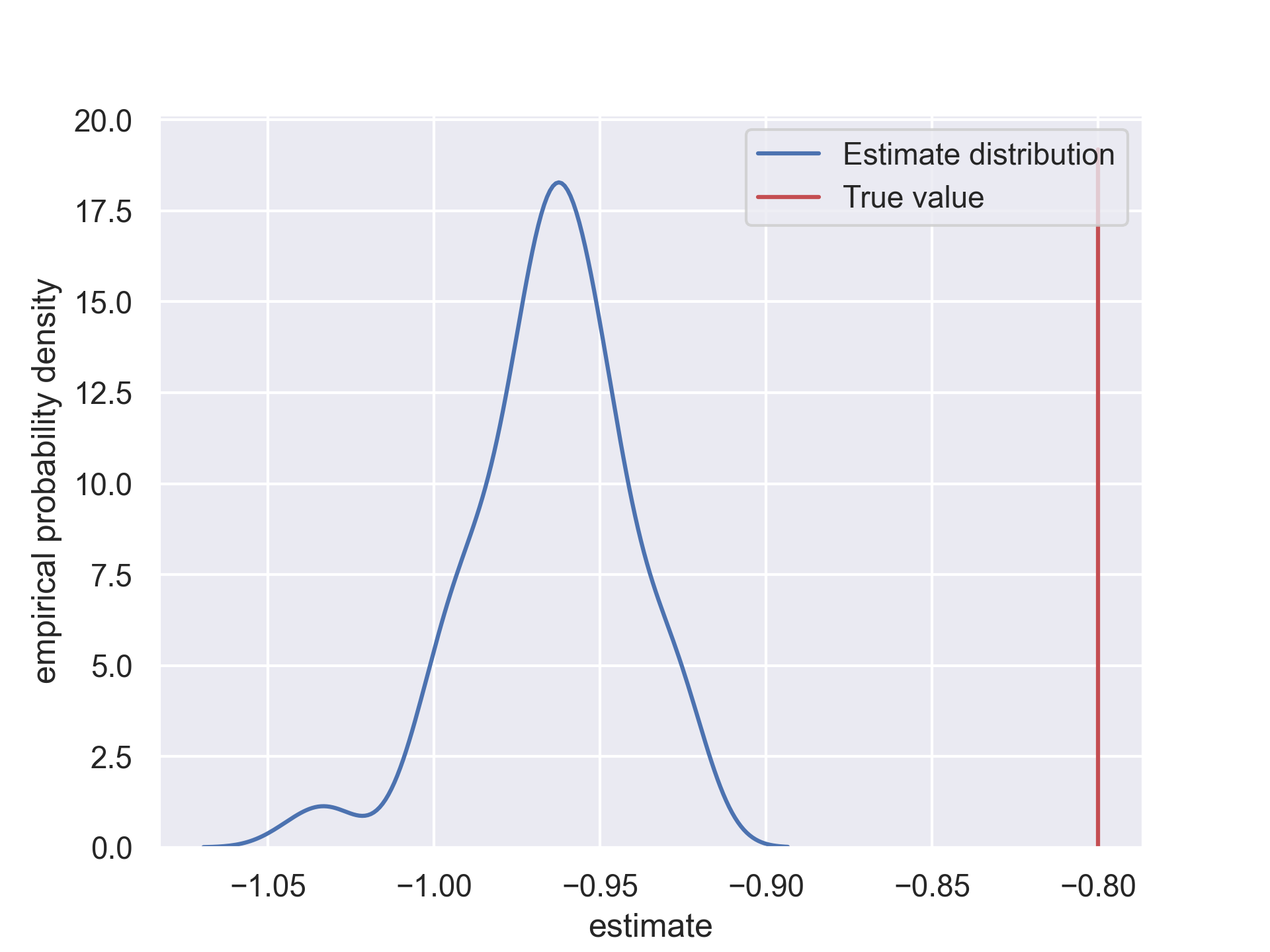}
         \caption{parameter $\mu_J$}
         \label{fig:exemplary_results_mu_J}
     \end{subfigure}
     \hfill
     \begin{subfigure}[b]{0.45\textwidth}
         \centering
         \includegraphics[width=\textwidth]{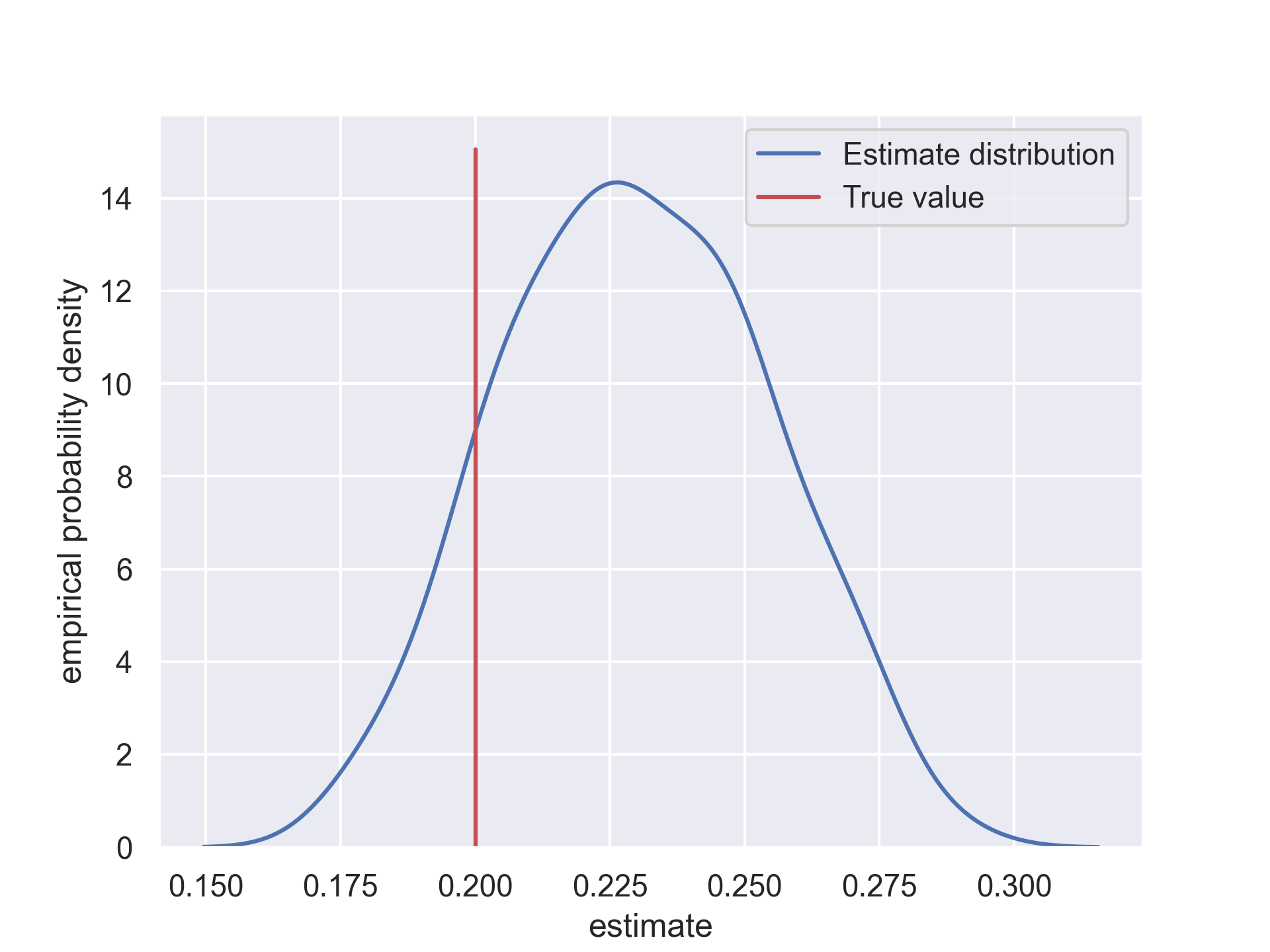}
         \caption{parameter $\sigma_J$}
         \label{fig:exemplary_results_sig_J}
     \end{subfigure}
        \caption{Empirical PDFs made of exemplary Heston parameter estimates.}
        \label{fig:exemplary_results}
\end{figure}

We present here an exemplary estimation of the Heston model with jumps, to show the outcomes of the entire procedure. We assumed relatively non-informative prior distributions, with expected values shifted form the true parameters to make the task more challenging for the algorithm and to better reflect the real-life situation in which priors used are most of the time not matching true parameters exactly, but should be rather close to them. Table \ref{tab:exemplary_priors} lists all the values of priors which we used. Table \ref{tab:exemplary_results} summarises the results obtained and Fig. \ref{fig:exemplary_results} elaborates on those results by showing empirical distributions of samples for all parameters of the model. 

Analysing estimate samples for each of the parameters (presented in Fig. \ref{fig:exemplary_results}) one can observe that for most of them (Figs. \ref{fig:exemplary_results_mu}--\ref{fig:exemplary_results_rho} and \ref{fig:exemplary_results_sig_J}) the true value of the parameter is within the support of the distribution of all samples. However, in case of two parameters~---~$\lambda$ and $\mu_J$ (Figs.~\ref{fig:exemplary_results_lambda} and \ref{fig:exemplary_results_mu_J} respectively)~---~the scope of samples generated by the estimation procedure seems not even to include the parameter's true value. This is due to the fact that those parameters are related to intensity and size of jumps and for the simulation parameters which we picked jumps do not happen frequently (same as in case of real-life price falls). Hence, despite the procedure correctly identifies the moments of jumps and estimates their sizes, those estimates are relatively far from the true values simply because there was very little source material for the estimation in the first place. To be precise --- the stock price simulated for our exemplary estimation experienced four jumps, and times of those jumps have been easily identified by our procedure with almost 100\% certainty. Thus, since the length of time of the price observation (in years) was $T=3$, the most probable value of the jump intensity $\lambda$ was around $\frac{4}{3}$ (compare to the actual result in Table \ref{tab:exemplary_results}), although, obviously, other values of $\lambda$ (slightly smaller or bigger) could have also lead to four jumps and this is exactly what happened in our case, as our true intensity was $\lambda=1$ (again, see Table \ref{tab:exemplary_results}). Similarly, in case of $\mu_J$, the reason for the estimated average jump to be bigger (in terms of magnitude) than the actual one was that the four jumps which were simulated all happened to be more severe than the true value of $\mu_J$ would suggest (by pure chance) and this "pushed" the procedure towards overestimating the (absolute) size of the jump. 

\begin{table}[h]
\centering
\begin{tabular}{c|c} 
    \textbf{Prior parameter} & \textbf{Value} \\[0.5ex] 
    \hline\hline
    $\mu_0^\eta$ & 1.00125 \\
    $\sigma_0^\eta$ & 0.001 \\
    \hline
    $\bm{\Lambda}_0$ & $\begin{bmatrix}
                                10 & 0 \\
                                0 & 5
                        \end{bmatrix}$ \\
    $\bm{\mu}_0$ & $\begin{bmatrix}
                        35\cdot10^{-6} \\
                        0.988
                    \end{bmatrix}$ \\
    \hline        
    $a_0^\sigma$ & 149 \\
    $b_0^\sigma$ & 0.025 \\
    \hline        
    $\mu_0^\psi$   & $-0.45$ \\
    $\sigma_0^\psi$ & 0.3 \\
    $a_0^\omega$ & 1.03 \\
    $b_0^\omega$ & 0.05 \\
    \hline        
    $\lambda^{th}$ & 0.15 \\
    $\mu_0^J$ & $-0.96$ \\
    $\sigma_0^J$ & 0.3 \\
\end{tabular}
\caption{Priors for the exemplary estimation procedure}
\label{tab:exemplary_priors}
\end{table}

\begin{table}[h]
\begin{tabular}{c|c|c|c} 
    \textbf{Parameter} & \textbf{True value} & \textbf{Estimated value} & \textbf{Relative Error [\%]} \\[0.5ex] 
    \hline\hline
    $\mu$ & 0.1 & 0.09829 & 1.77 \\
    \hline
    $\kappa$ & 1 & 1.2190 & 21.90 \\
    \hline
    $\theta$ & 0.05 & 0.0493 & 1.92 \\
    \hline
    $\sigma$ & 0.01 & 0.0108 & 8.55 \\
    \hline
    $\rho$ & -0.5 & 0.4379 & 12.40 \\
    \hline
    $\lambda$ & 1 & 1.3349 & 33.49 \\
    \hline
    $\mu_J$ & -0.8 & -0.9651 & 20.64 \\
    \hline
    $\sigma_J$ & 0.2 & 0.2298 & 14.88 \\
\end{tabular}
\caption{Results of the exemplary estimation procedure.}
\label{tab:exemplary_results}
\end{table}

\subsection{Important findings}
\label{sec:important_findings}
Although estimation through the joint forces of Bayesian inference, Monte Carlo Markov Chains and particle filtering is generally considered very effective~~\cite{johannes_chapter_2010}, it has several areas the user needs to be aware of while using this estimation scheme. One of the issues worth considering is the impact of the prior parameters. Bayesian estimator of any kind needs to be fed with parameters of the prior distribution which should reflect our preexisting beliefs of what the value of the actual estimated parameter could be. The amount of information conveyed by a prior can be different, depending on several factors. One of them are parameters of the prior distribution itself. Consider $\mu^\eta_0$ and $\sigma^\eta_0$, mentioned already in the previous section. They are the prior parameters for $\eta$ - the predecessor for the $\mu$ estimates. The bigger $\sigma^\eta_0$ we take, the more volatile the estimates of $\eta$  --- and hence, $\mu$ --- are going to be. This is a pretty intuitive fact, being a direct consequence of the Bayesian approach itself. A more subtle influence of priors is hidden in the alternation between the MCMC sampling and particle filtering procedures. 

\begin{figure}
     \centering
     \begin{subfigure}[b]{0.45\textwidth}
         \centering
         \includegraphics[width=\textwidth]{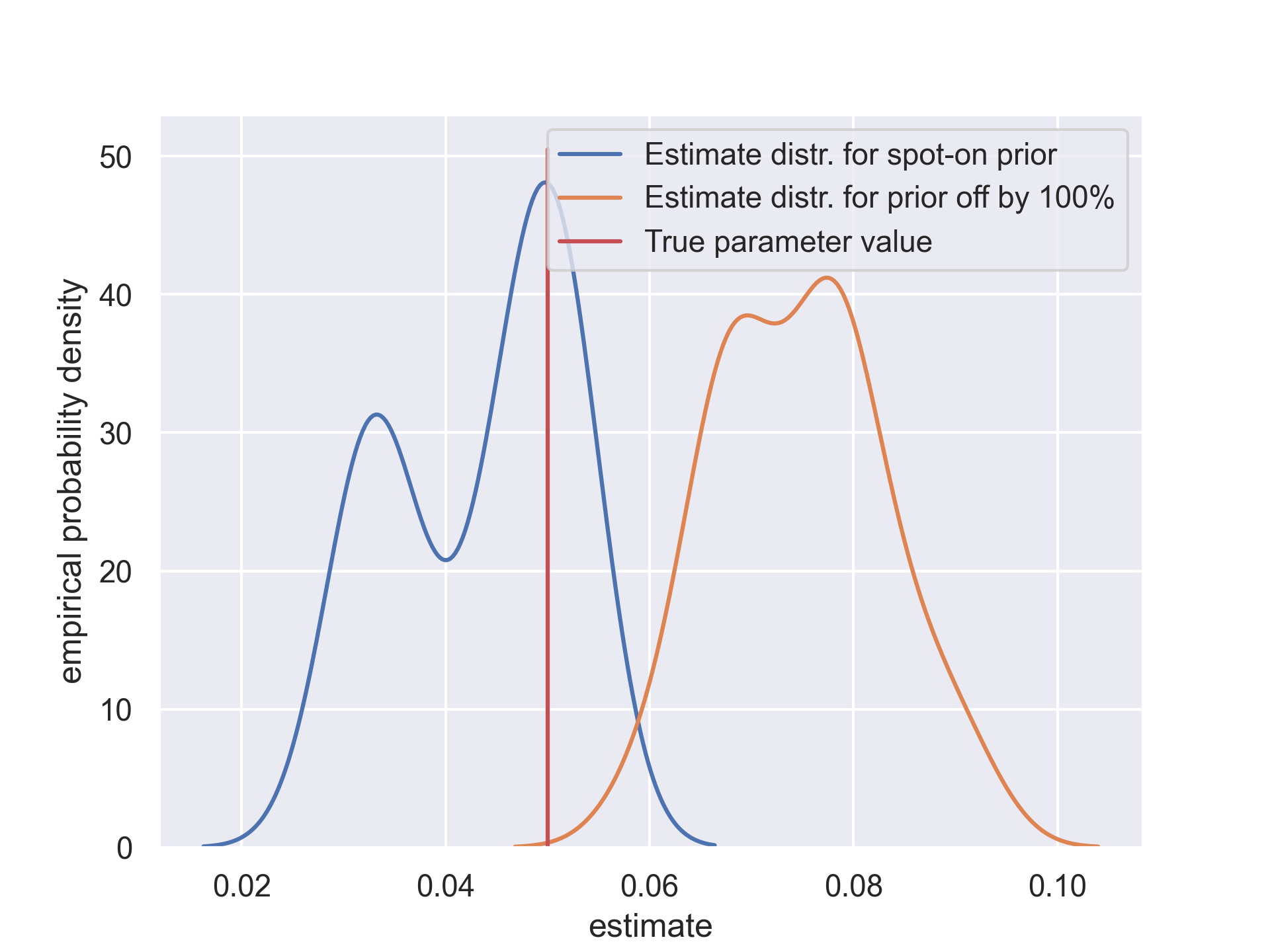}
         \caption{}
         \label{fig:mc_impact_low}
     \end{subfigure}
     \hfill
     \begin{subfigure}[b]{0.45\textwidth}
         \centering
         \includegraphics[width=\textwidth]{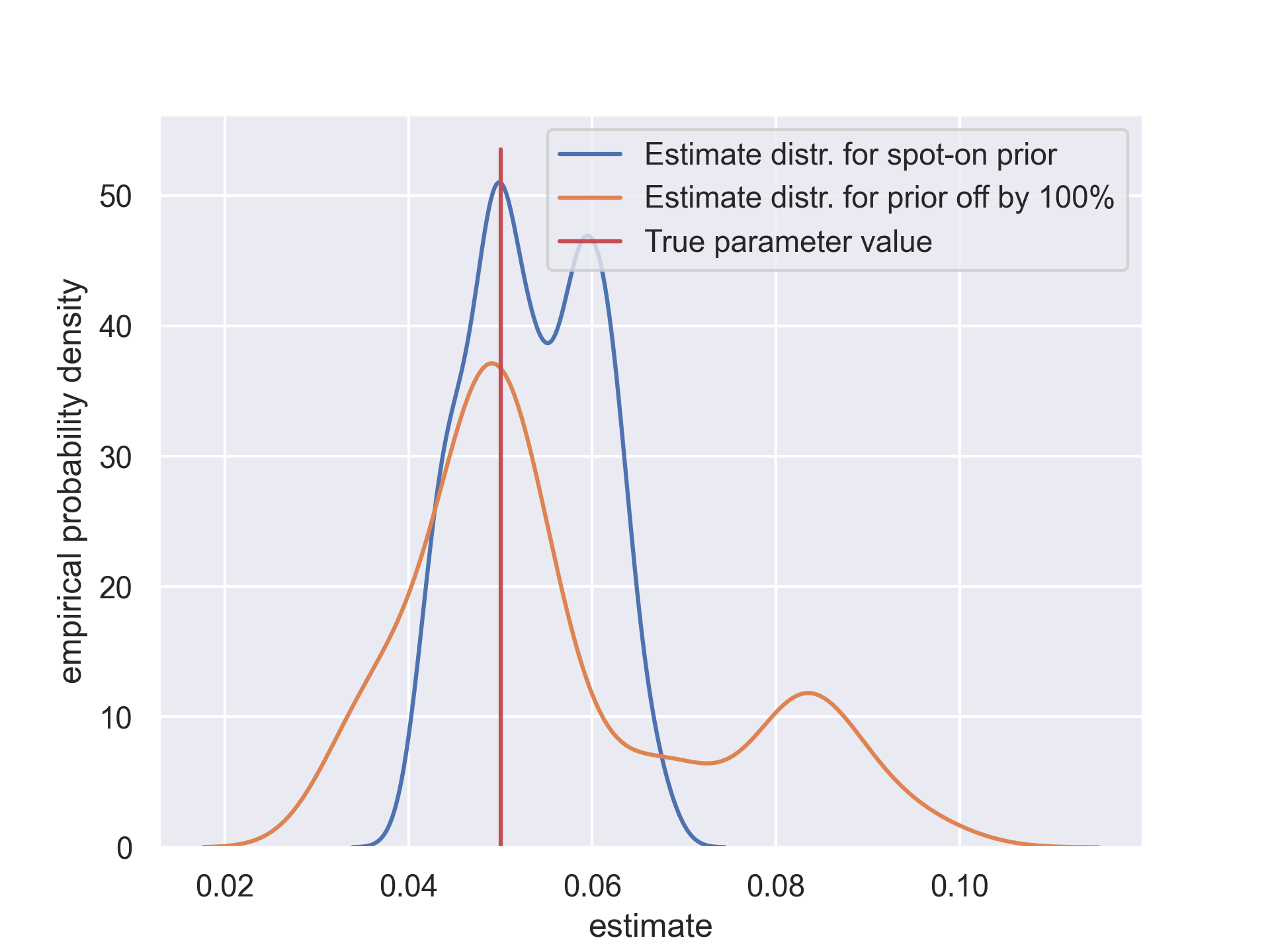}
         \caption{}
         \label{fig:mc_impact_high}
     \end{subfigure}
        \caption{Empirical distributions of the estimate samples for the $\theta$ parameter in case when the mean of the prior distribution matches exactly the true parameter and when it is twice bigger. Distributions in figure \ref{fig:mc_impact_low} was based on 10 sampling cycles, and the one in figure \ref{fig:mc_impact_high} --- 500 cycles. One can observe that for the first figure the distribution with spot-on prior gives very good results, much better then the shifted one. In the second figure, both distributions are comparable.}
        \label{fig:mc_impact}
\end{figure}

\begin{figure}[ht]
\centering
\includegraphics[width=\textwidth]{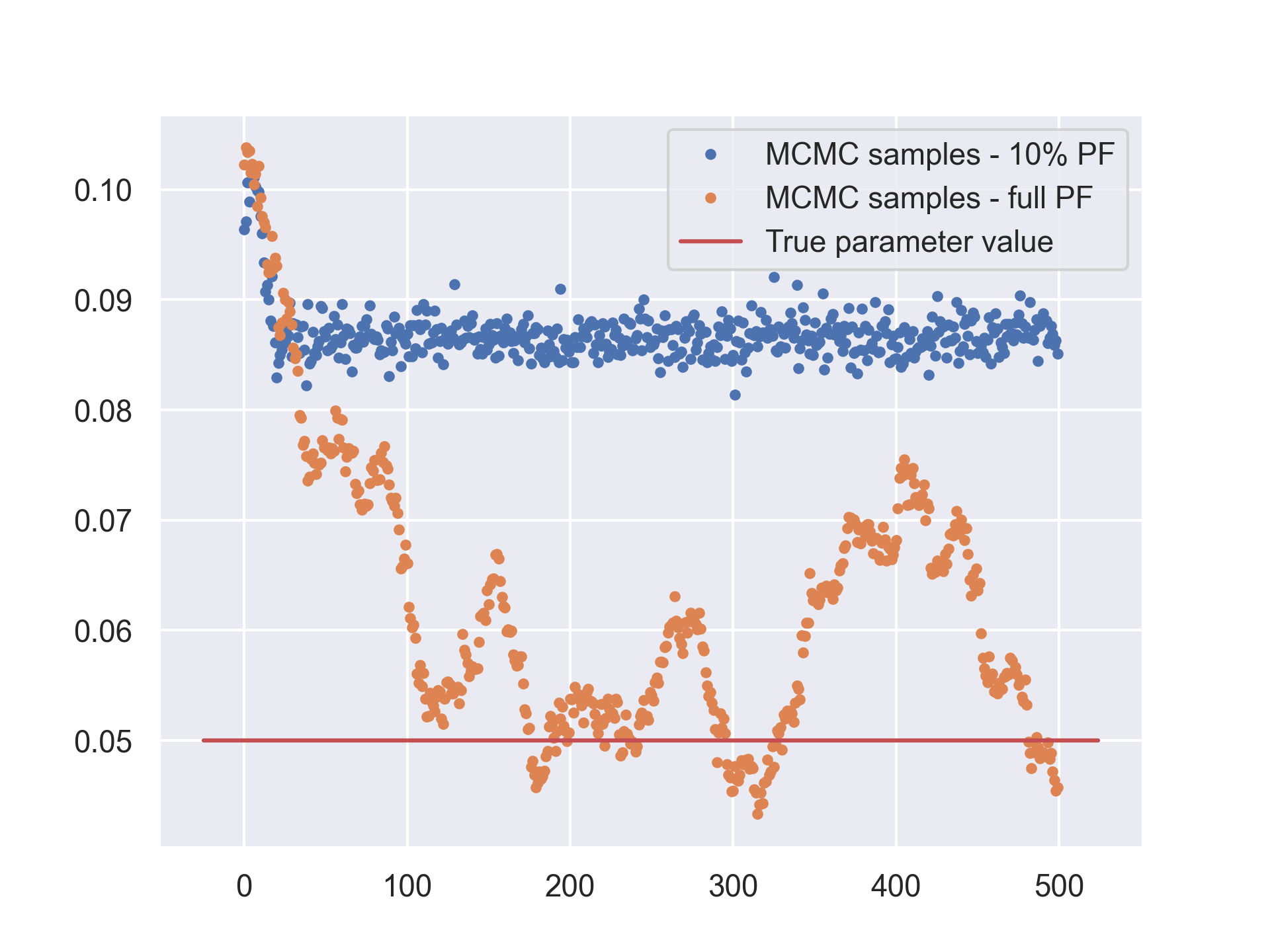}
\caption{Sequences of estimate samples for a procedure in which particle filtering was done only for the first 5\% samplings and another one, in which particle filtering was done for all the samplings. In both cases the mean of the prior distribution was shifted by 100\% compared to its true value. It can be observed that the samples of the first procedure get stuck around the value close to the one dictated by the prior, whereas samples of the other procedure converge to the true value of the parameter, which leads to the better final result, less dependent on the prior parameters.}
\label{fig:pf_impact}
\end{figure}

As mentioned in the previous section MCMC and particle filtering depend on one another. As can be seen in the Algorithms \ref{alg:est_Heston} and \ref{alg:est_Heston_jumps}, we have taken the approach that the particle filtering procedure should be done first and singular parameters which it needs should be the expected values of the prior distributions which we assume. Having volatility estimated that way, we can estimate the parameters, then based on them re-estimate the volatility process and so on. Although we can keep alternating that way as many times as we want, till the planned end of the estimation procedure, one might be tempted to perform the particle filtering procedure fewer times, as it is much more computationally expensive than the MCMC draws. The premise for that would be that after several trials, the volatility estimate becomes "good enough" and from that point onward, one can only generate more MCMC samples. A critical observation that we have made is that the quality of the initial volatility estimates depends very highly on the prior parameters which were used to initiate it. With little number of particle filtering procedures followed by multiple MCMC draws, the entire scheme does not have "enough time" to properly calibrate and results tend to stick to the priors which have been used. That means --- for a prior leading exactly to the true value of the parameter --- the estimator returns almost error-less results, however, if one uses a prior leading to value of the true parameter, e.g. 20\% bigger than it really is --- the estimate will probably be off by roughly 20\%, which does not make the estimator very useful. A counter-proposal can then be made, to perform particle filtering as long as possible. This however, is not an ideal solution either. Firstly, as we said, it is very computationally expensive, and secondly --- a very long chain of samples increases the probability that the estimation procedure would at some point return an "outlier", i.e. an estimate really far away from the true value of the parameter, which is especially likely if we use meta-parameters responsible for such parameter's variance (like e.g. $\sigma_0^\eta$ for $\eta$) bigger. Appearance of such "outliers" is especially unfavourable in case of the MCMC methods, since its nature is that each sample is directly dependent on the previous one, so the whole procedure is likely to "stay" in the given "region" of the parameter space for a some number of subsequent simulations, thus impacting the final estimate of the parameter (which is the mean of all observed samples). Therefore --- a clear trade-off appears. If one believes strongly that the prior they use is rather correct and only needs some "tweaking" to adjust it to the particular data-set --- a modest number of particle filtering can be applied\footnote{for applications in finance, this task is sometimes easier than for some other fields of science, as numerous works have been published already, presenting the results of the estimates of well-known stocks or market indices within various models --- see e.g. Ref~\cite{eraker_impact_2003}}, followed by an arbitrary number of MCMC draws. If however we do not know much about our data-set and do not want to convey too much information through the prior --- even at the cost of a bit worse final results --- they should run particle filtering bigger number of times. The visual interpretation of this rule has been presented in~Figs. \ref{fig:mc_impact} and \ref{fig:pf_impact}.

\begin{figure}[ht]
\centering
\includegraphics[width=\textwidth]{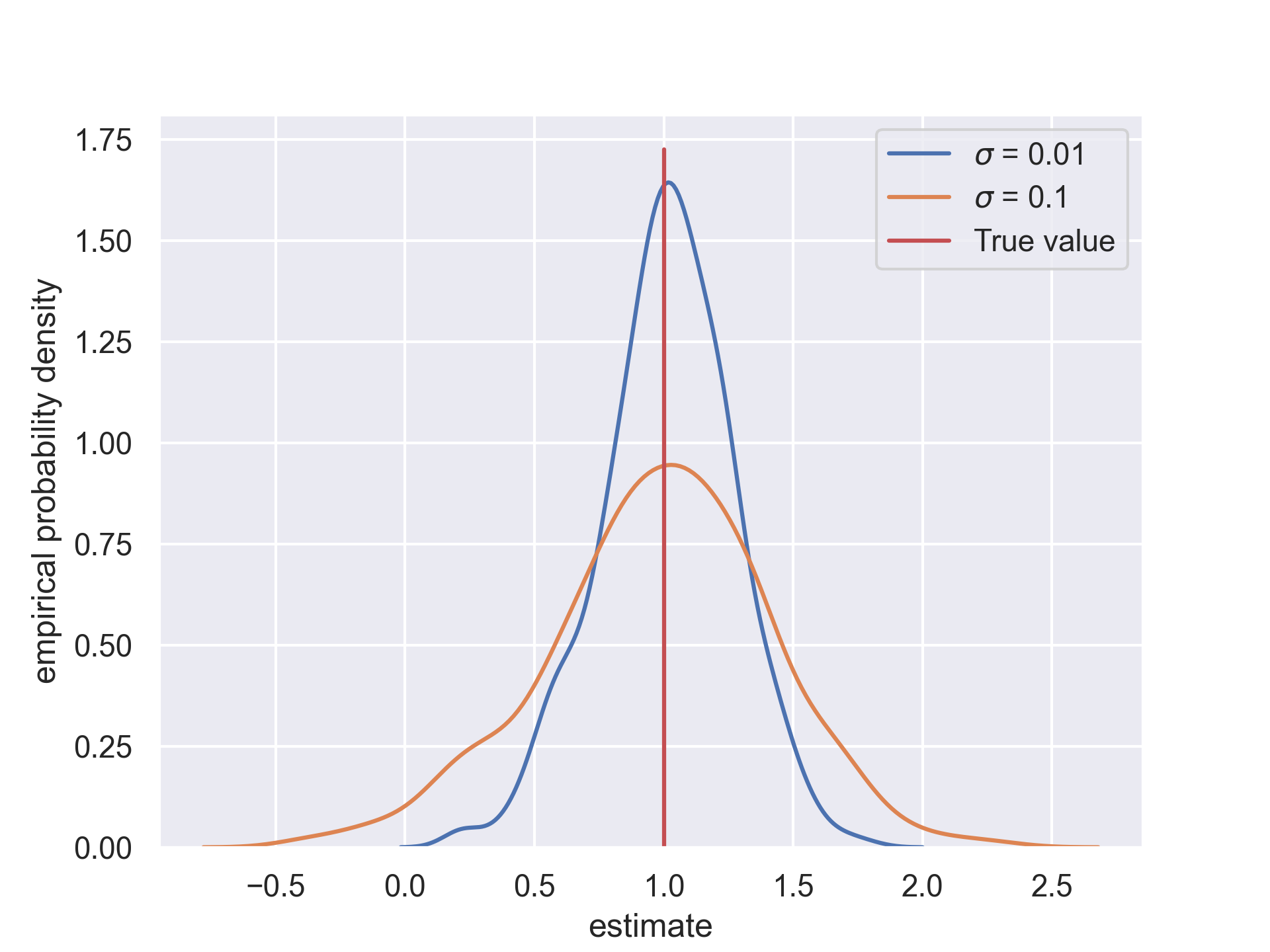}
\caption{Empirical distributions of the estimate samples for the $\kappa$ parameter of two different trajectories of the Heston model - one with $\sigma=0.01$ and the other for $\sigma=0.1$. The distribution of the estimate samples of the trajectory with smaller value of $\sigma$ is narrower and more concentrated, hence --- is likely to give less variable final estimates.}
\label{fig:other_param_impact}
\end{figure}

Another factor which should be taken into consideration using the Bayesian approach for estimating Heston model is that the quality of results depends highly on the very parameters we try to estimate. The $\sigma$ parameter seems to play a critical role for the Heston model in particular. This can be observed in Fig. \ref{fig:other_param_impact}. To produce it, an identical estimation procedure has been performed for two sample paths (which we can think of as of two different stocks). They have been simulated with the very same parameters, besides $\sigma$. Path no. 1 has been simulated with $\sigma = 0.01$ and path no. 2 with $\sigma = 0.1$, i.a. ten times bigger. The histograms present the distribution of the estimated values of the $\kappa$ parameter, true $\kappa$ was $\kappa=1$ and the red vertical line illustrates this true value. It is clearly visible, that for the value of $\sigma=0.01$ the samples were much more concentrated around the true value, while for a bigger value $\sigma=0.1$ --- they are more dispersed and the variance of the distribution is significantly bigger. This incommodity cannot be easily resolved, as the true values of parameters of the trajectories are idiosyncratic --- they cannot be influenced by the estimation procedure itself. However, we wanted to sensitise the reader to the fact that the bigger the value of $\sigma$, the less trustworthy the results of the estimation of the other parameters might be.

\subsection{Towards real-life applications}
At the end we wanted to emphasise on the applicability of the methods described above to the real market data. We briefly mentioned in the Introduction that results yield by some investment strategies are dependent on the character of the asset, which might be accurately captured by parameters of the models we use to model this asset (see Ref.~\cite{gruszka_advanced_2021} for more details). Having a tool for obtaining the values of those parameters from the data opens up a new field of investigation relying on comparing synthetically generated stock price trajectories to the real ones and selecting optimal portfolio management strategies based on the results of such comparisons. 

\section{Conclusions}
\label{sec:Conclusions}

In this paper, a complete estimation procedure of the Heston model without and with jumps in the asset prices is presented. Bayesian regression combined with the particle filtering method is used as the estimation framework. Although some parts of the procedure have been already used in the past, our work gives the first complete follow-along recipe of how to estimate the Heston model to the real stock market data. Moreover, we presented a novel approach to handle jumps in order to neutralize their negative impact on the estimates of the key parameters of the model. And we proposed an improvement of the sampling in the particle filtering method to get a better estimate of the volatility.

We extensively analysed the impact of prior parameters as well as the number of MCMC samplings and particle filtering iterations on the performance of our procedure. Our findings may help to avoid several difficulties related with the Bayesian methods and to apply them successfully to the estimation of the model.

Our results have an important practical impact. In one of our recent papers \cite{gruszka_advanced_2021} we have shown that the relative performance of several investment strategies within the Heston model varies with the values of its parameters. In other words, what turned out to be the best strategy in one range of the parameter values may the worst one in the others. Thus, figuring out which parameter of the model correspond to a given stock market will allow one to choose the optimal investment strategy for that market.

\appendix
\section{Table of symbols}
\label{app:table_of_symbols}

\begin{longtable}{l|p{10cm}}
Quantity & Explanation \\ \hline\hline
$T$ & max time (i.e. $t\in[0, T]$)\\
$S(t)$ &  asset price (with $S(0)\equiv S_0$)\\
$v(t)$ &  volatility (with $v(0)\equiv v_0$)\\
$B^S(t)$ &  Brownian motion for the price process\\
$B^v(t)$ &  Brownian motion for the volatility process\\
$\mu$ &  drift\\
$\kappa$ &  rate of return to the long-time average\\
$\theta$ &  long-time average\\
$\sigma$ &  volatility of the volatility\\
$\rho$ &  correlation between prices and volatility\\
$Z(t)$ &  size of the jump\\
$\mu^J$ &  mean of the jump size\\
$\sigma^J$ &  standard deviation of the jump size\\
$q(t)$ &  Poisson process counting jumps\\
$\lambda$ &  intensity of jumps\\
$\Delta t$ & time step (aka discretisation constant)\\
$n$ &  number of time steps (aka length of data)\\
$\varepsilon^S(t)$ & price process random component \\
$\varepsilon^v(t)$ & volatility process random component\\
$\varepsilon^{add}(t)$ &  additional random component -- see eq. \eqref{eq:eps_v})\\
$\eta$ & regression parameter for drift estimation -- see eq. \eqref{eq:eta}\\
$R(t)$ & ratio between neighbouring prices -- see eqs. \eqref{eq:Returns}, \eqref{eq:Returns_jumps}\\
$y^S(t)$ &  series of dependent variables for the drift estimation -- see eq. \eqref{eq:y_s}\\
$x^S(t)$ &  series of independent variables for the drift estimation -- see eq. \eqref{eq:x_s}\\
$\mathbf{y}^S$ &  vector of the dependent variable for the drift estimation -- see eq. \eqref{eq:y_s_vec}\\
$\mathbf{x}^S$ &  vector of the independent variable for the drift estimation -- see eq. \eqref{eq:x_s_vec}\\
$\mu_0^\eta$ &  mean of the prior distribution of $\eta$\\
$\sigma_0^\eta$ &  standard deviation of the prior distribution of $\eta$\\
$\tau_0^\eta$ & precision of the prior distribution of $\eta$\\
$\hat{\eta}$ & OLS estimator of $\eta$ -- see \eqref{eq:eta_hat}\\
$\mu^\eta$ &  mean of the posterior distribution of $\eta$ -- see \eqref{eq:mu_eta} \\
$\tau^\eta$ &precision of the posterior distribution of $\eta$ -- see \eqref{eq:tau_eta}\\
$\eta_i$ & $i$-th sample of $\eta$ -- see \eqref{eq:eta_i_bayes}\\
$\mu_i$ & $i$-th estimate of the drift -- see \eqref{eq:mu_i}\\
$\beta_1$ & regression parameter for volatility parameters estimation -- see eq. \eqref{eq:beta_1}\\
$\beta_2$ & regression parameter for volatility parameters estimation -- see eq. \eqref{eq:beta_2}\\
$\bm{\beta}$ & vector of regression parameters for volatility parameters estimation -- see eq. \eqref{eq:beta_vec}\\
$\mathbf{y}^v$ &  vector of the dependent variable for the volatility parameter estimation -- see eq. \eqref{eq:y_v_vec}\\
$\mathbf{x}_1^v$ &  vector of the independent variable for the volatility parameter estimation -- see eq. \eqref{eq:x_1_v_vec}\\
$\mathbf{x}_2^v$ &  vector of the independent variable for the volatility parameter estimation -- see eq. \eqref{eq:x_2_v_vec}\\
$\mathbf{X}^v$ &  matrix of the independent variable for the volatility parameter estimation -- see eq. \eqref{eq:x_v_vec}\\
$\bm{\varepsilon}^v$ & noise vector of the volatility parameter estimation -- see eq. \eqref{eq:eps_v_vec}\\
$\bm{\mu}_0^\beta$ & mean vector of the prior distribution of $\bm{\beta}$\\
$\bm{\Lambda}_0^\beta$ & precision matrix of the prior distribution of $\bm{\beta}$\\
$\bm{\mu}^\beta$ & mean vector of the posterior distribution of $\bm{\beta}$ -- see eq. \eqref{eq:mu_beta}\\
$\bm{\Lambda}^\beta$ & precision matrix of the posterior distribution of $\bm{\beta}$ -- see \eqref{eq:Lambda_beta}\\
$\hat{\bm{\beta}}$ & OLS estimator of $\bm{\beta}$ -- see \eqref{eq:beta_hat}\\
$\bm{\beta}_i$ & $i$-th sample of $\bm{\beta}$ -- see \eqref{eq:beta_i_bayes}\\
$\kappa_i$ & $i$-th estimate of $\kappa$ -- see \eqref{eq:kappa_i_bayes}\\
$\theta_i$ & $i$-th estimate of $\theta$ -- see \eqref{eq:theta_i_bayes}\\
$a_0^\sigma$ &  shape parameter of the prior distribution of $\sigma^2$\\
$b_0^\sigma$ &  scale  parameter of the prior distribution of $\sigma^2$\\
$a^\sigma$ &  shape parameter of the posterior distribution of $\sigma^2$\\
$b^\sigma$ &  scale parameter of the posterior distribution of $\sigma^2$\\
$\sigma_i$ & $i$-th estimate of the $\sigma$ -- see \eqref{eq:sigma_i_bayes}\\
$e_1^\rho(t)$ & series of residuals of the price equation -- see \eqref{eq:e1_rho}\\
$e_2^\rho(t)$ & series of residuals of the volatility equation -- see \eqref{eq:e2_rho}\\
$\psi$ & regression parameter for $\rho$ estimation, $\psi = \sigma\rho$ -- see \eqref{eq:rho_reg}\\
$\omega$ & regression parameter for $\rho$ estimation, $\omega = \sigma^2(1\rho^2)$ -- see \eqref{eq:rho_reg}\\
$e_1^\rho(t)$ & series of independent variables for the estimation of $\rho$ -- see \eqref{eq:e1_rho}\\
$e_2^\rho(t)$ & series of dependent variables for the estimation of $\rho$ -- see \eqref{eq:e2_rho}\\
$\mathbf{e}_1^\rho$ & vector of the independent variables for the estimation of $\rho$ -- see \eqref{eq:e1_rho_vec}\\
$\mathbf{e}_2^\rho$ & vector of the dependent variables for the estimation of $\rho$ -- see \eqref{eq:e2_rho_vec}\\
$\mathbf{e}^\rho$ & matrix of residuals -- see \eqref{eq:e_rho}\\
$\mathbf{A}^{\rho}$ & auxiliary matrix for solving $\rho$ regression -- see \eqref{eq:A_rho}\\
$\mu_0^\psi$ &  mean of the prior distribution of $\psi$\\
$\tau_0^\psi$ & precision of the prior distribution of $\psi$\\
$\mu^\psi$ &  mean of the posterior distribution of $\psi$ -- see \eqref{eq:mu_psi} \\
$\tau^\psi$ & precision of the posterior distribution of $\psi$ -- see \eqref{eq:tau_psi}\\
$a_0^\omega$ &  shape parameter of the prior distribution of $\omega$\\
$b_0^\omega$ &  scale  parameter of the prior distribution of $\omega$\\
$a^\omega$ &  shape parameter of the posterior distribution of $\omega$ -- see \eqref{eq:a_psi}\\
$b^\omega$ &  scale parameter of the posterior distribution of $\omega$ -- see \eqref{eq:b_psi}\\
$\psi_i$ & $i$-th sample of $\psi$ -- see \eqref{eq:psi_i_bayes}\\
$\omega_i$ & $i$-th sample of $\omega$ -- see \eqref{eq:omega_i_bayes}\\
$\varepsilon_j(t)$ & $j$-th sample of particle filtering independent errors -- see \eqref{eq:eps_pf}\\
$z_j(t)$ & $j$-th sample of particle filtering residuals -- see \eqref{eq:z_eps_pf}\\
$w_j(t)$ & $j$-th sample of particle filtering correlated -- see \eqref{eq:z_eps_pf}\\
$\widetilde{V}_j(t)$ & $j$-th raw volatility particle -- see \eqref{eq:V_tilde_pf}\\
$\widetilde{W}_j(t)$ & $j$-th particle likelihood measure -- see \eqref{eq:W_tilde_pf} and \eqref{eq:W_tilde_pf_jump}\\
$\breve{W}_j(t)$ & $j$-th particle probability -- see \eqref{eq:W_bow_pf}\\
$\mathbf{U}_j(t)$ & $j$-th particle-probability vector -- see \eqref{eq:U_pf}\\
$\widetilde{V}_j^{sort}(t)$ & $j$-th raw volatility particle in a sorted sequence -- see \eqref{eq:V_sorted_1_pf}, \eqref{eq:V_sorted_2_pf} and \eqref{eq:V_sorted_3_pf}\\
$\breve{W}_j^{sort}(t)$ & probability of the $j$-th particle in the sorted sequence -- see \eqref{eq:W_bow_sorted_pf}\\
$F_{\widetilde{V}^{sort}}(v)$ & continuous CDF of resampled particles  -- see \eqref{eq:F_V_pf}\\
$V_j(t)$ & $j$-th final volatility particle -- see \eqref{eq:V_pf}\\
$\lambda^{th}$ & proportion of particles encoding a jump\\
$\widetilde{J}_j(t)$ & $j$-th raw moment-of-a-jump particle, see \eqref{eq:J_tilde_pf}\\
$\mu_0^J$ & mean of the raw size-of-a-jump particles\\
$\sigma_0^J$ & standard deviation of the raw size-of-a-jump particle\\
$\widetilde{Z}_j(t)$ & $j$-th raw size-of-a-jump particle, see \eqref{eq:Z_tilde_pf}\\
$Z_j(t)$ & $j$-th resampled size-of-a-jump particle -- see \eqref{eq:Z_pf}\\
$\lambda(t)$ & probability of a jump -- see \eqref{eq:lambda_k}\\
$\lambda_i$ & $i$-th estimate of $\lambda$ -- see \eqref{eq:lambda_i}\\
$Z(t)$ & estimate of an average size of a jump -- see \eqref{eq:Z_k_pf}\\
$\mu_i^J$ & $i$-th estimate of $\mu^J$ -- see \eqref{eq:mu_J_i}\\
$\sigma_i^J$ & $i$-th estimate of $\sigma^J$ -- see \eqref{eq:sigma_J_i}
\end{longtable}

\bibliography{Bibliography}
\bibliographystyle{unsrt}

\end{document}